\documentclass[pra,twocolumn,showpacs,amsmath,amssymb,pra,floatfix]{revtex4}

\usepackage{graphicx}
\usepackage{dcolumn}
\usepackage{bm}

\begin{document}


\title{Lithium in strong magnetic fields}
\date{\today}

\author{Omar-Alexander Al-Hujaj}
\email{Alexander.Al-Hujaj@pci.uni-heidelberg.de}
\affiliation{%
Theoretische Chemie, Institut f\"ur Physikalische Chemie der Universit\"at Heidelberg, INF 229, 69120 Heidelberg, Germany}%
\author{Peter Schmelcher}
\email{Peter.Schmelcher@pci.uni-heidelberg.de}
\affiliation{%
Theoretische Chemie, Institut f\"ur Physikalische Chemie der Universit\"at Heidelberg, INF 229, 69120 Heidelberg, Germany}%
\affiliation{%
Physikalisches Institut der Universit\"at Heidelberg, Philosophenweg 12, 69120 Heidelberg, Germany}%

\date{\today}

\begin{abstract}
The electronic structure of the lithium atom in a strong magnetic field
$0 \le \gamma \le 10$ is investigated. Our computational approach is a full
configuration interaction method based on a set of anisotropic Gaussian orbitals
that is nonlinearly optimized for each field strength. Accurate results for 
the total energies and one-electron ionization energies for the ground and
several excited states for each of the symmetries  
$^20^{+}$, $^2(-1)^{+}$, $^4(-1)^{+}$, $^4(-1)^{-}$, $^2(-2)^{+}$,
$^4(-2)^{+}$, $^4(-3)^{+}$
are presented. The behaviour of these
energies as a function of the field strength is discussed and classified.
Transition wave lengths for linear and circular polarized transitions are
presented as well.

\end{abstract}

\maketitle
\section{Introduction}
During the past twenty years an enormous development
of our knowledge on atoms exposed to strong magnetic fields
has taken place (see the reviews
\cite{Friedrich:1989_1,Ruder:1994_1,Cederbaum:1997_1,Schmelcher:1998_1,Herlach:2003_1}
and references therein). Focusing on astrophysical conditions and on the field regime $100  \le B \le 10^5$~T for magnetic
white dwarfs it is in particular the one and two-electron problems, i.e. the hydrogen and helium atom,
whose behavior and properties in strong magnetic fields have been investigated in detail.
In both cases our knowledge on the electronic structure of the atoms has had major impact
on astrophysical observation. For the hydrogen atom a huge amount of data is nowadays available
both with respect to the bound state energy levels and transition moments \cite{Ruder:1994_1} as well as for the
continuum properties \cite{Merani:1995_1}. Among others, the corresponding data have lead to a conclusive
interpretation of the observed spectrum of the white dwarf GrW+70$^\circ$$8247$
which was a key to our understanding of the properties of spectra of magnetic white dwarfs in general
(see
e.g.
Refs.
\cite{Angel:1985_1,Angel:1978_1,Greenstein:1985_1,Wunner:1985_1,Wickramasinghe:1988_1}).

In the late nineties a powerful computational approach was developed and implemented in order to study many-electron
atomic problems in the presence of a strong magnetic field. During the past six years this approach
was applied in order to investigate the electronic structure of the helium atom thereby covering the 
complete regime of astrophysically relevant field strengths
\cite{Becken:1998_all,Al-Hujaj:2003_1}. Approximately 90 excited electronic states are now known with high accuracy thereby yielding $12000$ transition
wavelengths. As a consequence strong evidence arose that the mysterious
absorption edges of the magnetic white dwarf GD229 \cite{Green:1980_1,Schmidt:1990_1,Schmidt:1996_1}, which were for almost
25 years unexplained, are due to helium in a strong magnetic field $B
\approx 50\ 000$~T \cite{Jordan:1998_1,Jordan:2001_1}.
Also very recently the newly established helium data were used to analyze a number of
magnetic and suspected-magnetic southern white dwarfs \cite{Schmidt:2001_1,Wickramasinghe:2000_1}.

Although our knowledge on the electronic structure of hydrogen and helium in a strong magnetic field have
allowed for the interpretation of absorption features of a variety of magnetic
white dwarfs, there are other magnetic objects whose spectra can not be explained in terms of these species.
In addition, due to the increasing availability of observatories with higher
resolutions and sensitivities, new spectra have been obtained that
remain unexplained \cite{Reimers:1998_1}, thereby opening the necessity for studies of heavier atoms exposed to magnetic fields:
The ongoing Sloan Digital Sky Survey already doubled the number of
known magnetic white dwarfs \cite{Schmidtxx}. It is believed that these heavy atoms are present in the
atmospheres of the corresponding stars due to  accretion of interstellar
matter, and particularly it is expected that these objects are quite
common \cite{Reid:2001_1}. In spite of this interest in multi-electron atoms in strong magnetic fields there is only a very
scarce literature. One reason for this is certainly the conceptual and computational difficulties associated with
the competing electron-electron, electron-nuclear-attraction, paramagnetic and diamagnetic interactions which are
of comparable strength under astrophysical conditions.

The present work makes a start to fill the above-mentioned gap and develops a full configuration 
interaction (full CI) approach for multi-electron systems thereby focusing on the lithium atom in a strong
magnetic field. Let us comment at this point on the state-of-the-art of the literature on the
lithium atom exposed to the field thereby following a chronological order.
In Ref.~\cite{Neuhauser:1987_1} a combination of an adiabatic and Hartree-Fock (HF) approach
is employed to obtain ground state energies for four different field strengths in the high field regime.
Ref.~\cite{Demeur:1994_1} provides also values of the ground state energy via a HF adiabatic approach
in the high field regime. Ref.~\cite{Jones:1996_1} equally employs an unrestricted HF approach in order
to obtain the energies of the ground states of the symmetry subspaces $^20^{+},^2(-1)^{+},^4(-1)^{+},^4(-1)^{-}$
for the weak to intermediate regime of field strengths
$\gamma=0-5$ ($\gamma$ denotes the magnetic field strength in
atomic units, where $\gamma=1$ correspondes to $2.355\ 10^5$~T). Ref.~\cite{Ivanov:1998_1} contains also a HF
investigation of the $1^20^{+}$, $1^2(-1)^{+}$, $1^4(-1)^{+}$,
$1^2(-1)^{-}$, $1^4(-3)^{+}$ electronic states in the
complete regime $\gamma=0-1000$. Neutral atoms for nuclear charge
numbers $Z=1-10$ in the high field 
regime are investigated in Ref.~\cite{Ivanov:2000_1} equally within a HF approach. The crossovers of the symmetries
of the ground states are discussed and analyzed in detail. Ref.~\cite{Qiao:2000_1} uses a so-called frozen-core
approach to simplify the three-electron problem in a strong magnetic field. This was the first fully correlated
approach to the lithium atom although it is only valid, i.e. reliable, for not too strong magnetic fields.
The three energetically lowest states of $^20^{+}$, $^2(-1)^{+}$, $^2(-2)^{+}$ symmetry have been studied
for the regime $\gamma=0-5.4$. More recently \cite{Guan:2001_1} adds
to these results  a denser grid of 
field strengths for the same regime of field strengths and provides
also a few oscillator strengths of 
the corresponding transitions. Finally \cite{Mori:2002_1} provides
some results on the ground state 
energies of neutral atoms $Z=1-26$ for a few field strengths. 

The present investigation goes in several respects significantly
beyond the results in the existing literature 
on lithium in a strong magnetic field. First of all it covers the complete weak to intermediate regime of
field strengths $\gamma=0-10$ and more importantly we provide accurate results of the energies and transition
wave lengths for many excited states that have not been studied so far
employing a fully correlated approach. The ground and many excited states
for each of the symmetry subspaces $^20^{+}$, $^2(-1)^{+}$,
$^4(-1)^{+}$, $^4(-1)^{-}$, $^2(-2)^{+}$, $^4(-2)^{+}$, $^4(-3)^{+}$
are investigated thereby yielding a total of 28
states and their behavior as a 
function of the field strength for a grid of $11$ field strengths in
the above-mentioned regime. This  
multiplies the existing knowledge on the electronic structure of the
lithium atom in strong fields. 

In detail we proceed as follows. Section \ref{sec:hamil} provides the
electronic Hamiltonian and discusses its 
symmetries. Section  \ref{sec:method} contains a description of our
full configuration interaction approach and its 
implementation as well as remarks on the basis set of nonlinearly
optimized anisotropic Gaussian orbitals. 
Section \ref{sec:res}, which is the central part of this work,
presents the results i.e. the total and ionization  
energies for the ground and many excited states for a variety of
symmetries. Section \ref{sec:wave} yields the wavelengths
of the electromagnetic transitions. We close with a  summary in
section \ref{sec:concl}.

\section{Hamiltonian and symmetries}
\label{sec:hamil}

The starting point of our investigations is the  electronic
Hamiltonian
 for infinite nuclear mass, which takes in atomic units
 (a.u.) and for the symmetric gauge of the vector potential the
 following form: 
\begin{eqnarray}
  \label{eq:ham}
  H(\bm{B})&=&\sum_{i=1}^3 H_i(\bm{B}) + \frac{1}{2}\sum_{i\not=j}H_{ij}\qquad \mbox{with}\\
  H_i(\bm{B})&=&\frac{\bm{p}_i^2}{2}+\frac{\bm{B}\cdot\bm{l_i}}{2}+\frac{(\bm{B}\times\bm{r}_i)^2}{8}-\frac{3}{|\bm{r}_i|}+\frac{g
  \bm{B}\cdot\bm{s}_i}{2}\\
H_{ij}&=&\frac{1}{|\bm{r}_i-\bm{r}_j|}
\end{eqnarray}
Here $H_i(\bm{B})$ represents the one-particle  Hamiltonian of the
$i$-th particle and $H_{ij}$ is the
two-particle interaction between particles $i$ and $j$. Specifically $H_i(\bm{B})$ contains
the Zeeman-term $1/2\ \bm{B}\cdot\bm{l_i}$, which represents the
interaction of the magnetic field with the angular momentum of the
electron, the diamagnetic term $1/8(\bm{B}\times\bm{r}_i)^2$, the
Coulomb interaction with the nuclear charge $-3/|\bm{r}_i|$, and
the spin Zeeman-term $g/2\  \bm{B}\cdot\bm{s}_i$.  The two-particle
operators represents the electron-electron Coulomb repulsion.

If the magnetic field is chosen to point in $z$~direction, the
component of the total angular momentum along the $z$-axis $M$, the total spin $S$, the $z$~projection of the
total spin $S_z$, and the total $z$~parity $\Pi_z$ are conserved. In the
following we use the spectroscopic notation $\nu^{2S+1}M^{\Pi_z}$ for
the electronic states. Here
$\nu$ stands for the degree of excitation, with respect to the
specified symmetry. In the following all total energies are given for
the spin beeing  maximal
polarized antiparallel to the direction of the magnetic field
(i.e. $S_z=-S$). Energies for other spin projections can be obtained
by adding the corresponding spin Zeeman-energy difference.

\section{Numerical method}
\label{sec:method}

The Schr\"{o}dinger equation is solved by applying a full CI
approach. The basic ingredient is 
an anisotropic Gaussian basis set, which was put forward by Schmelcher
and Cederbaum \cite{Schmelcher:1988_1}, and  
which has been
successfully applied to several atoms, ions and
molecules \cite{Kappes:1996_1,Detmer:1998_1,Becken:1998_all,Al-Hujaj:2003_1}.
The corresponding basis functions have been optimized for each field
strength, and each symmetry separately, in order to solve   different
one- and two-particle 
problems, i.e. H, 
Li$^+$, and 
Li$^{2+}$, in a magnetic field of the corresponding strength. Therefore a
nonlinear optimization procedure has been applied, which has been worked
out in our group (see  Refs.~\cite{Becken:1998_all,Al-Hujaj:2003_1}).

Our lithium calculations have been performed using a configurational
basis set of three-electron  Slater determinants. The latter are
constructed from the canonical orthogonal one-particle basis set (see Ref.~\cite{Szabo_Ostlund}), which is
obtained by the following cut-off technique. In the first step
the overlap matrix $\bm{S}(m_j,\pi_{z_j})$  of the primitive Gaussian
orbitals is constructed. Its eigenvectors
$\{\bm{v}_{s_j}(m_j,\pi_{z_j})\}$ and 
corresponding eigenvalues $\{e_{s_j}(m_j,\pi_{z_j})\}$ are determined. For the
following calculations we restrict the number of eigenvectors
$\bm{v}_{s_j}$, to those which posses eigenvalues $e_{s_j}$ above an
appropriate chosen threshold $\varepsilon$. This way we avoid
quasi-linear dependencies in the configuration space generated by our
optimized basis set.
With the remaining vectors $\bm{v}_{s_k}$ the Schr\"{o}dinger equation for the
one-particle Hamiltonian
$H_i(\bm{B})$ is mapped on an ordinary matrix  eigenvalue problem. The latter
is solved 
numerically and the resulting eigenvectors
$\{\bm{h}_i(m_j,\pi_{z_j})\}$ serve as the spatial
part of our one-particle basis set for the electronic structure
calculations.  The spinors
$\chi_j$ are products 
of  this orthogonal
one-particle 
basis functions 
$\bm{h}_i(m_j,\pi_{z_j})$ and the usual spin 
eigenfunctions $\bm{\alpha}$ or $\bm{\beta}$. Three-electron Slater
determinants  are  
constructed from spinors obeying the correct symmetries, i.e. 
\begin{eqnarray}
  m_1+m_2+m_3&=&M \\
  \pi_{z_1}\pi_{z_2}\pi_{z_3}&=&\Pi_z\\
  s_{z_1}+s_{z_2}+s_{z_3}&=&S_z.
\end{eqnarray}

In order to keep the one-particle basis set as small as possible,  an
appropriate  selection scheme 
for the basis functions is crucial. This concerns the selection of
the symmetries of the one-particle functions as well as the selection from  appropriate
sets of orbitals, which result from the above mentioned nonlinear
optimizations. 

In general the core electrons of doublet states of the lithium atom,
i.e. the  $1s^2$ configuration, are well described by 
functions optimized for the Li$^+$ $1^10^+$ state. Therefore we
applied a two-particle optimization procedure to
functions with the one-particle symmetries $m^{\pi_z}=0^+$ and
$m^{\pi_z}=0^-$. Further orbitals involved in the calculation of the
doublet states of the lithium atom are obtained by
 optimizing orbitals for the  hydrogen atom 
 $Z=1$.

For the fully spin-polarized quartet states, the
situation is different. Electrons are much less correlated and
therefore the computationally demanding two-particle optimizations are not truely necessary. 
 Core electrons, i.e. $m^{\pi_z}=0^+$
and $m^{\pi_z}=0^-$ are 
 described by functions optimized for Li$^{2+}$, energetically higher
 orbitals such as
$m^{\pi_z}=(\pm1)^+$, $m^{\pi_z}=(\pm1)^-$ 
are taken from basis sets optimized for $Z=2$, others from basis
sets optimized for $Z=1$.

Typically the one-particle basis sets consist of approximately 100
Gaussian functions, which give rise to $8\,000$~--~$40\,000$
three-electron slater-determinants, depending on the addressed
symmetry subspace. Very sophisticated algorithms allow to 
calculate the full Hamiltonian matrix efficiently. We exploit the
fact, that  the Hamiltonian matrix is a sparse and symmetric
one and apply a Lanczos algorithm for its diagonalization.

\section{Total and ionization energies}
\label{sec:res}

\subsection{Total energies and global ground states}
\label{sec:glob_ground}

The symmetries of the global ground states of individual atoms or ions
change depending on the field strength
\cite{Ivanov:1998_1,Ivanov:1999_1,Ivanov:2000_1,Ivanov:2001_1,Ivanov:2001_2}:  In different field regimes
eigenstates with different symmetries represent the ground state of
the system. Therefore the global ground state of an atom or ion
experiences a series of 
crossovers. These crossovers emerge to the delicate interplay between
the different terms of the Hamiltonian in the field such as the spin
Zeeman, diamagnetic and Coulomb interaction. Of particular importance
are here the magnetically tightly bound orbitals that represent a key
ingredient of strongly bound atomic or ionic states in strong
fields. The number of ground state crossovers for a certain atom or
ion in the field cannot be predicted e.g. on the grounds of symmetry
reasoning but has to determined through electronic structure
calculations in the presence of the field. The above holds in
particular for the lithium atom considered here.

\begin{figure}[htbp]
  \centering
  \includegraphics*[scale=0.3]{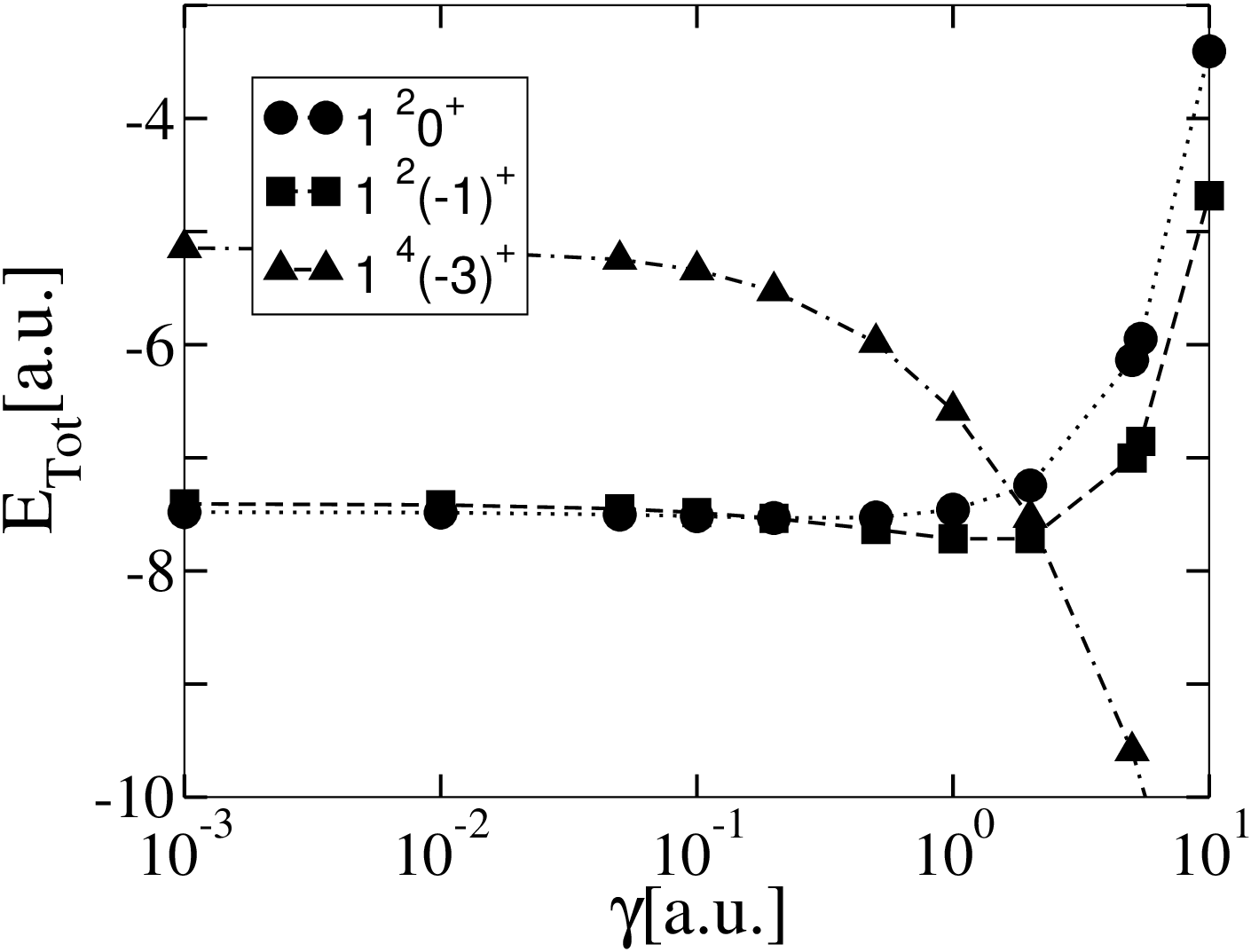}
  \caption{Total energies $_{\mbox{\scriptsize Tot}}$ of the global
  ground states of the lithium atom in a.u. as function of the magnetic field
  strength $\gamma$.} 
  \label{fig:groundstate_li}
\end{figure}

 The total
energies of the components of the global electronic ground state for
lithium as a 
function of the field strength are depicted in
Fig.~\ref{fig:groundstate_li}.  
For $\gamma=0$ and in the low field regime
$0< \gamma \leq 0.1929$ the state $1^20^+$ represents the ground
state. It is a doubly tightly bound state,
i.e. it involves two  tightly bound orbitals of $1s$ character
(although we are employing full CI calculations here, we will
occasionally use the mean-field (HF) orbital notation to
elucidate the character of the fully correlated atomic wave function). For
 doublet states ($S_z=-1/2$), the total energy decreases for weak
fields, due to the spin Zeeman-term, and it increases for strong
fields, which is  a consequence of the predominance of the increasing
(positive definite)  kinetic energy. For the $1^20^+$ state the total energy
passes through a 
minimum at $\gamma \approx0.304$~a.u.
In the
intermediate field regime ($0.1929\leq\gamma\leq2.210$) the ground
state of the lithium atom is represented by the triply tightly bound
state $1^2(-1)^+$, which contains in particular the dominant
$1s^22p_{-1}$ configuration.
 Fig.~\ref{fig:groundstate_li}  shows, that the total energy of this
state also passes through a minimum, which is at higher  field
strengths ($\gamma\approx 1.466$), compared to the position of the
minimum of the low field ground
state. The total energy of the triply tightly bound quartet state
$1^4(-3)^+$, which 
represents the 
ground state of the lithium atom for high field strengths ($\gamma>2.210$), is
dominated by the spin Zeeman-term ($S_z=-3/2)$. This results in a
monotonously 
decreasing total energy.

Our values for the field strengths corresponding to the crossovers of
the global ground state deviate about $10\%$ 
from previously published values for the first crossover and
approximately 3\% for the second crossover. Ivanov (HF
calculations) \cite{Ivanov:1998_1} 
 states for the first crossing field $\gamma$ $0.17633$ and Guan
 (modified full core plus correlation method)
$0.1753$ \cite{Guan:2001_1}. The field strength belonging to the
second crossover is found by Ivanov at
$\gamma=2.153$ compared to our value of $2.210$.  

\subsection{Ionization threshold}
\label{sec:ionisation}

In order to calculate one-particle binding energies the ionization
threshold has to be defined.
In the following, we will define
one-particle ionization as a process, that brings one electron  to
infinity  and thereby conserving
all quantum numbers of the
atomic state. The one-particle ionization threshold $E_T(M,S_z)$
for a state with magnetic quantum number $M$ and $z$~projection $S_z$ of the
total spin   is defined in the following way:
\begin{equation}
  E_T(M,S_z)=\min_{M_1,S_{z_1}} E^{\mbox{\scriptsize
  Li}^+}(M_1,S_{z_1})+E^{\mbox{\scriptsize e}^-}(M_2,S_{z_2})
\end{equation}
where $E^{\mbox{\scriptsize Li}^+}(M_1,S_{z_1})$ and
$E^{\mbox{\scriptsize e}^-}(M_2,S_{z_2})$ are the total energies of
the Li$^+$ ion and the electron, respectively, depending on their magnetic
quantum numbers $M_i$ and $z$~projection $S_{z_i}$
($i=1,2$) of the total spin. The quantum numbers for the electron $M_2$ and $S_{z_2}$
can be expressed in terms of the ionic and the atomic quantum numbers
\begin{equation}
  M_2=M-M_1 \qquad S_{z_2}=S_z-S_{z_1}.
\end{equation}
This procedure has to be repeated for each symmetry and field
strength in order to identify the corresponding  threshold. Therefore 
several energy levels of Li$^+$ have to be considered as a function of
the field strength. Table \ref{tab:li+} shows the total energies for
the  Li$^+$ states, associated to one-particle ionization thresholds.

\begin{table}[ht]
\begin{ruledtabular}
\begin{tabular}{d*{3}{d}}

\multicolumn{1}{c}{\rule[-0.7em]{0mm}{2.0em}} &\multicolumn{1}{c}{$1^10^+$} & \multicolumn{1}{c}{$1^30^+$} & \multicolumn{1}{c}{$1^3(-1)^+$} \\
\hline
\multicolumn{1}{c}{\rule[-0.7em]{0mm}{2.0em}$\gamma$ } &
\multicolumn{1}{c}{$E_{\mbox{\scriptsize tot}}$ [a.u.]}
&\multicolumn{1}{c}{$E_{\mbox{\scriptsize{}tot}}$ [a.u.]}&
\multicolumn{1}{c}{$E_{\mbox{\scriptsize{}tot}}$ [a.u.]}\\
\hline
 0.000  & -7.277191  & -5.110633  &  -5.026321    \\
 0.001  & -7.277189  & -5.111640  &  -5.027815    \\
 0.010  & -7.277327  & -5.120614  &  -5.041247    \\
 0.020  & -7.277376  & -5.110313  &  -5.056040    \\
 0.050  & -7.277336  & -5.159107  &  -5.099595    \\
 0.100  & -7.276897  & -5.204480  &  -5.169539    \\
 0.200  & -7.274673  & -5.286753  &  -5.300455    \\
 0.500  & -7.259522  & -5.483980  &  -5.643726    \\
 1.000  & -7.205547  & -5.727321  &  -6.119216    \\
 2.000  & -7.004453  & -6.126974  &  -6.899768    \\
 5.000  & -5.891947  & -7.170075  &  -8.636273    \\
 5.400  & -5.704147  & -7.292325  &  -8.827671    \\
10.000  & -3.153453  & -8.490652  &  -10.659060   \\
\end{tabular}
\caption{Total energies for Li$^+$ associated to one-particle
  ionization thresholds at different field strengths for the considered
  lithium states.\label{tab:li+}}
\end{ruledtabular}
\end{table}
 
In the following, we will present our results for the total energies
and the one-particle
ionization energies for a variety of states of the lithium atom with
different symmetries. 

\subsection{The symmetry subspace $^20^+$}
\label{sec:ionisation_res}

\begin{figure}[htbp]
  \centering
  \includegraphics*[scale=0.30]{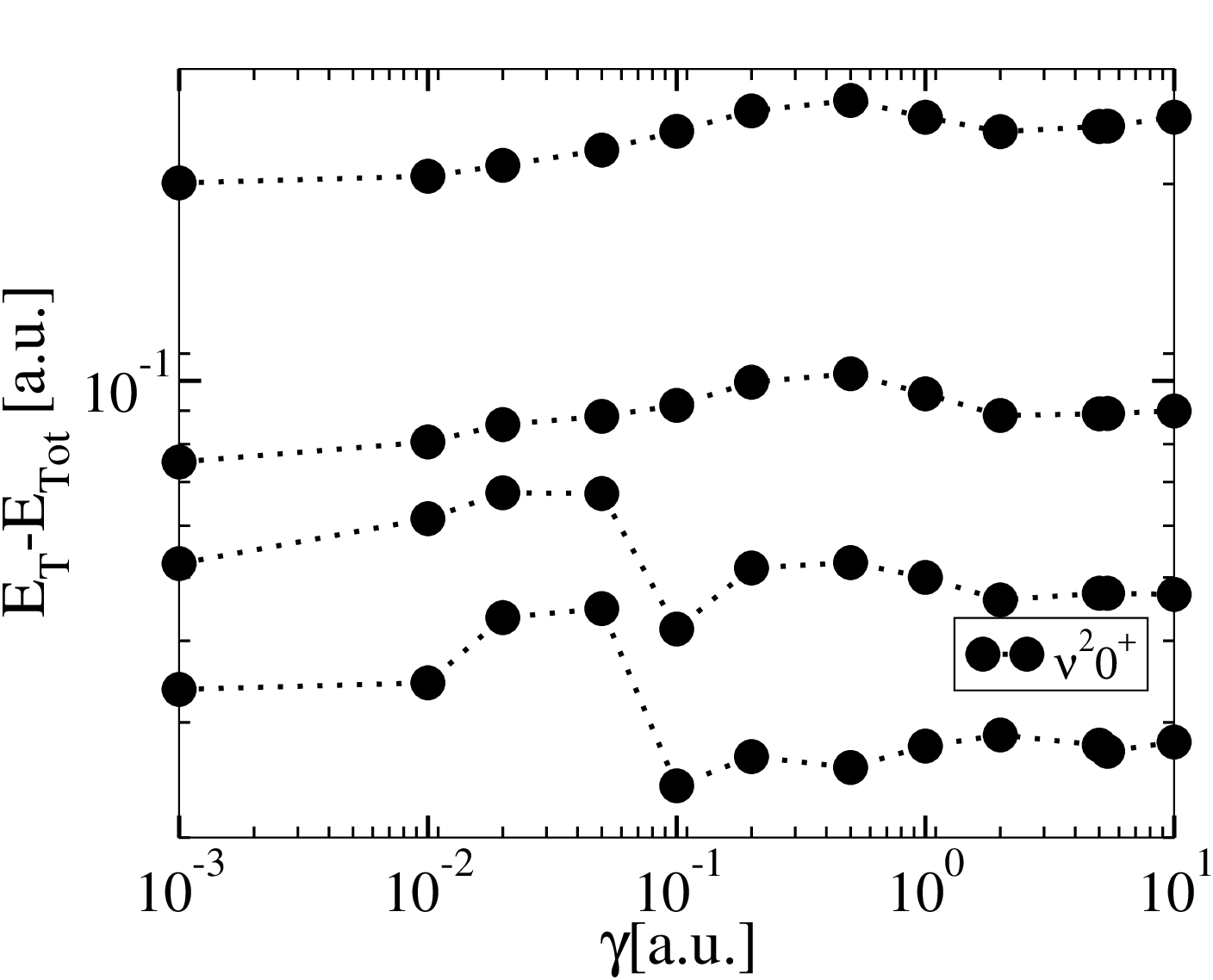}
  \caption{One-particle ionization energies for the states $\nu^20^+$ ($\nu=1,2,3,4$)  as a function of the magnetic field strength $\gamma$.}
  \label{fig:li_m0d+}
\end{figure}

We present in Fig.~\ref{fig:li_m0d+} the one-particle ionization
energies for the $\nu^20^+$ states ($\nu=1,2,3,4$), and in
table~\ref{tab:n20+}  numerical values for the corresponding
total  
energies, one-particle ionization energies, including previously
published data for the total energies. The ionization threshold for
these and for all other considered doublet states, is associated with the
 Li$^+$ state $1^10^+$.
The energetically lowest of
the states in the $^{2}0^+$ symmetry subspace, represents as mentioned
above the global ground state of the 
atom for low magnetic field strengths. Comparing the total energies to the
previously published data, shows, that the relative accuracy for
$\gamma=0$ for the ground  state is  $4\cdot10^{-5}$, $5\cdot10^{-4}$ for the
state $2^20^+$, and $4\cdot10^{-3}$ for the $3^20^+$ state. For finite field strengths
our results are significantly below the Hartree-Fock results
\cite{Ivanov:1998_1,Jones:1996_1} and at least for $\gamma\ge0.1$ below the
correlated results of Guan \cite{Guan:2001_1}. 

In  Fig.~\ref{fig:li_m0d+} it can be seen, that the
 one-particle ionization energy of the ground state only weakly
 depends on the  
field strength. For a vanishing  field it amounts to $0.20058$~a.u.,
 whereas at 
a field strength of $\gamma=10$ it is $0.25310$~a.u.  It increases for
 weak to intermediate field strengths and possesses a maximum in the
 intermediate field regime.  A similar
statement holds for the first excited state of this symmetry
subspace, i.e. for the state $2^20^+$. The one-particle ionization
 energies for  the states
$3^20^+$ and $4^20^+$ exhibit a more 
pronounced  dependence on the field strength. This is especially true in the intermediate
field regime, where an avoided crossing between the $2^20^+$, $3^20^+$
 and $4^20^+$ occurs.
\squeezetable
\begin{table*}[ht]
\begin{ruledtabular}
\begin{tabular}{c*{3}{c}*{7}{c}}
 &\multicolumn{3}{c}{\rule[-0.7em]{0mm}{2.0em}{}$1^20^+$}& \multicolumn{2}{c}{$2 ^20^+$}& \multicolumn{2}{c}{$3 ^20^+$}& \multicolumn{2}{c}{$4 ^20^+$}\\
\hline
\multicolumn{1}{c}{\rule[-0.7em]{0mm}{2.0em}{}$\gamma$}&  \multicolumn{1}{c}{E$_{\mbox{\scriptsize{}tot}}$} &\multicolumn{1}{c}{E$_{\mbox{\scriptsize{}Ion}}$}  & \multicolumn{1}{c}{E$_{\mbox{\scriptsize{}Lit}}$}   & \multicolumn{1}{c}{E$_{\mbox{\scriptsize{}tot}}$} &\multicolumn{1}{c}{E$_{\mbox{\scriptsize{}Ion}}$}   & \multicolumn{1}{c}{E$_{\mbox{\scriptsize{}tot}}$} &\multicolumn{1}{c}{E$_{\mbox{\scriptsize{}Ion}}$} & \multicolumn{1}{c}{E$_{\mbox{\scriptsize{}tot}}$} &\multicolumn{1}{c}{E$_{\mbox{\scriptsize{}Ion}}$} \\
\hline
 0.000 & -7.477766  & 0.200575&-7.47806032310\footnotemark[1]& -7.350744  &0.073553     & -7.304474  &0.0272828      &  -7.280117 &0.002925\\
 0 (Lit) &            &         &                         &-7.354076\footnotemark[5]&& -7.3355235\footnotemark[6]&&  -7.318315\footnotemark[5]      &     \\
 0.001 & -7.478032  & 0.200843&-7.43326\footnotemark[2]      & -7.352286  &0.075097     & -7.329739  &0.052550 & -7.310861  &0.033672\\
 0.010 & -7.482888  & 0.205562&-7.43760\footnotemark[2]      & -7.357941  &0.080615     &  -7.338823 &0.061497 & -7.311831  &0.034505 \\ 
 0.020 & -7.490983  & 0.213607&-7.44214\footnotemark[2]      & -7.363118  &0.085743     & -7.344767  &0.067391 & -7.320737  &0.043361 \\ 
 0.050 & -7.502724  & 0.213607&                              & -7.365504  &0.088169     & -7.344529  &0.067193 & -7.322185  &0.044850 \\  
 0.100 & -7.517154  & 0.240838&-7.5137817\footnotemark[3]    & -7.367564  &0.090667     & -7.317517  &0.040620 & -7.300920  &0.024023 \\  
 0.200 & -7.533495  & 0.258822&-7.48400\footnotemark[2]      & -7.374189  &0.099516     & -7.326335  &0.051662 & -7.301269  &0.026596 \\ 
 0.500 & -7.528055  & 0.268532&-7.5235946\footnotemark[3]    & -7.361991  &0.102469     & -7.312259  &0.052736 & -7.285148  &0.025626 \\  
 1.000 & -7.458550  & 0.253003&-7.40879\footnotemark[2]      & -7.301070  &0.095523     & -7.255573  &0.050026 & -7.233152  &0.027605 \\  
 2.000 & -7.244919  & 0.240466&-7.19621\footnotemark[2]      & -7.092907  &0.088454     & -7.050638  &0.046185 & -7.033148  &0.028695 \\ 
 5.000 & -6.136918  & 0.244971&-6.08811\footnotemark[2]      & -5.980919  &0.088972     & -5.939235  &0.047289 & -5.919658  &0.027712 \\ 
 5.400 & -5.949297  & 0.245150&-5.8772\footnotemark[4]       & -5.793212  &0.089065     & -5.751426  &0.047279 & -5.731222  &0.027075 \\  
10.00 & -3.406556   & 0.253103&-3.35777\footnotemark[2]      & -3.243308  &0.089855     & -3.200544  &0.047091 & -3.181432  &0.027979\\  
\end{tabular}
\caption{Total energies E$_{\mbox{\scriptsize{}Tot}}$, one-particle ionization
  energies E$_{\mbox{\scriptsize{}Ion}}$ and previously published results for the
  total energies E$_{\mbox{\scriptsize Lit}}$ at different field strengths $\gamma$ in a.u. for the states $\nu ^20^+$ ($\nu=1,2,3,4$).}\label{tab:n20+}  \footnotetext[1]{Ref. \cite{Yan:1995_1}} \footnotetext[2]{Ref. \cite{Ivanov:1998_1}} \footnotetext[3]{Ref. \cite{Guan:2001_1}} \footnotetext[4]{Ref. \cite{Jones:1996_1}} \footnotetext[5]{Ref. \cite{King:1991_1}}\footnotetext[6]{Ref. \cite{Pipin:1992_1}}
\end{ruledtabular}
\end{table*}

\subsection{The symmetry subspace $\nu^2(-1)^+$}
\label{sec:n2-1+}

\begin{figure}[htbp]
  \includegraphics*[scale=0.30]{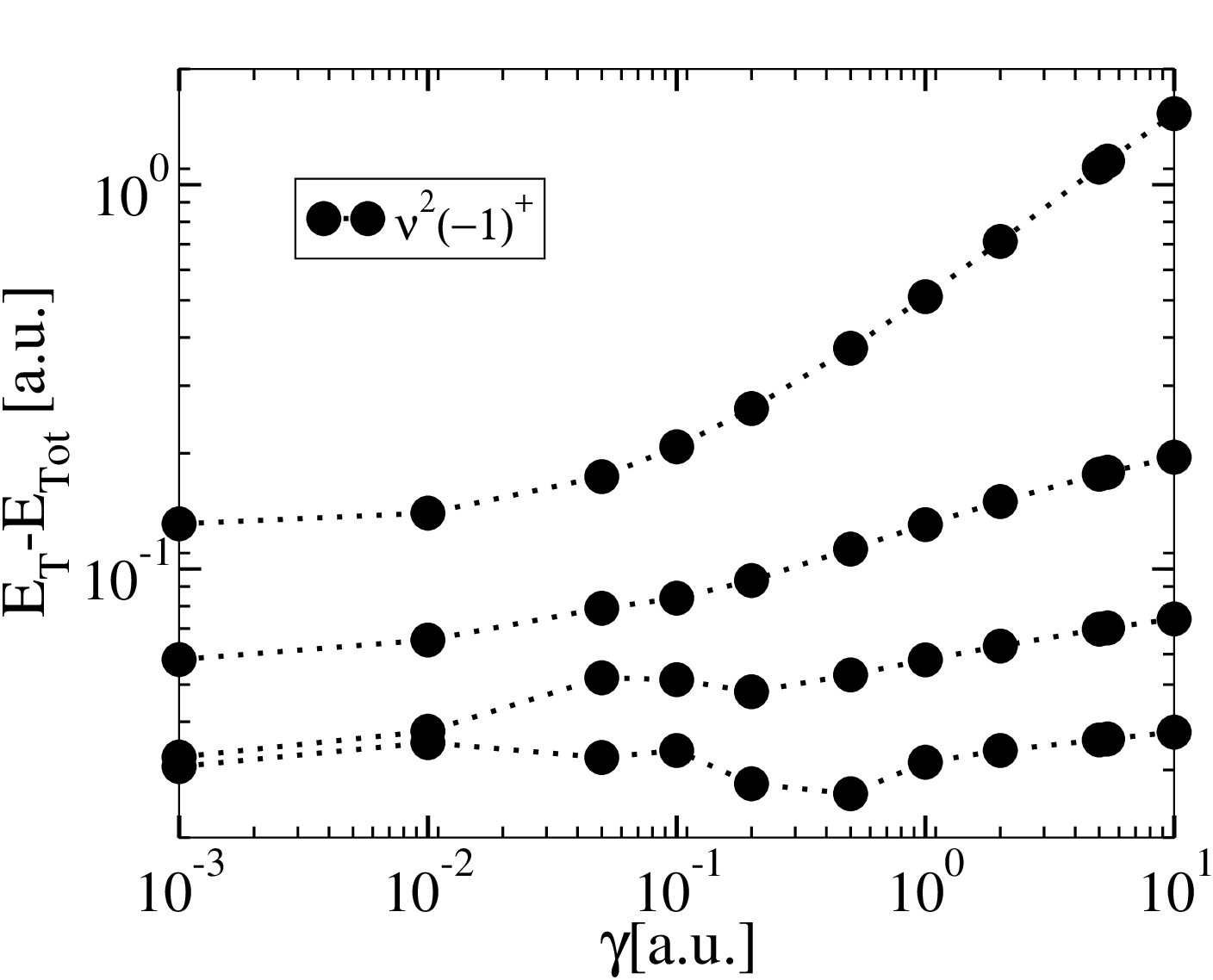}
  \caption{One-particle ionization energies for the states $\nu^2(-1)^+$ ($\nu=1,2,3,4$) in a.u. as a function of the magnetic field strength $\gamma$.}
  \label{fig:12-1+}
\end{figure}

The ground state of the symmetry subspace
$^2(-1)^+$ represents the global ground state of the lithium atom in the
intermediate field regime. It is a triply tightly bound
state beeing predominantly described by the orbitals $1s^22p_{-1}$. Therefore its one-particle ionization energy increases much
more rapidly with increasing field strength, than the corresponding
energy  of the low field ground state $1^20^+$. This is shown in
Fig.~\ref{fig:12-1+} and numerical values for the states $\nu^2(-1)^+$
($\nu=1,2,3,4$)  are listed in Tab.~\ref{tab:n2-1+}. Compared to the accurate,
zero field results of Sims \cite{Sims:1975_1}, it can be seen, that
the relative accuracy for the states $1^2(-1)^+$, $2^2(-1)^+$, and
$3^2(-1)^+$ is about $4\cdot10^{-4}$. For finite fields the comparison
of the total energies for
the $1^2(-1)^+$ state is as follows:
our energies are significantly below the Hf energies
\cite{Ivanov:1998_1},  for $\gamma<0.5$ above the correlated
results in Ref. \cite{Guan:2001_1} and for for $\gamma\ge0.5$ below them.

Furthermore the  table~\ref{tab:n2-1+} shows, that the state
$1^2(-1)^+$ becomes  for
$\gamma\ge0.2$ the most tightly bound state, i.e. the state with
the highest one-particle ionization energy. This holds even
for high fields, although there the quartet
states have a much lower total energy. For  the state $1^2(-1)^+$ the
one-particle ionization energy increases about more than one order of
magnitude in the considered field range:
 $0.129935$~a.u. at zero magnetic field, and $1.530623$~a.u. at
$\gamma=10$. 

The one-particle ionization energy of the first excited state
$2^2(-1)^+$ increases also monotonously as a
function of the field strength. At $\gamma=0$ it is $0.05581$ a.u., for
$\gamma=10$ it becomes $0.19525$~a.u. and therefore increases almost
by  a
factor of 4. For the higher excited states $3^2(-1)^+$ and $4^2(-1)^+$
the effect of an avoided crossing can be observd. Therefore the
one-particle ionization energies of these states show a more
complex behavior. As a result their ionization energy increases from
$\gamma=0$ to $\gamma=10$
to a much lower extent, than the ionization energy for the ground and the first excited state.

\begin{table*}[htpb]
\begin{ruledtabular}
\begin{tabular}{c*{3}{c}*{9}{c}}
 &\multicolumn{3}{c}{\rule[-0.7em]{0mm}{2.0em}{}$1^2(-1)^+$}& \multicolumn{3}{c}{$2 ^2(-1)^+$}& \multicolumn{3}{c}{$3 ^2(-1)^+$}& \multicolumn{2}{c}{$4 ^2(-1)^+$}\\
\hline
\multicolumn{1}{c}{\rule[-0.7em]{0mm}{2.0em}{}$\gamma $}&  \multicolumn{1}{c}{E$_{\mbox{\scriptsize{}tot}}$} &\multicolumn{1}{c}{E$_{\mbox{\scriptsize{}Ion}}$}  & \multicolumn{1}{c}{E$_{\mbox{\scriptsize{}Lit}}$}   & \multicolumn{1}{c}{E$_{\mbox{\scriptsize{}tot}}$} &\multicolumn{1}{c}{E$_{\mbox{\scriptsize{}Ion}}$} & \multicolumn{1}{c}{E$_{\mbox{\scriptsize{}Lit}}$} &  \multicolumn{1}{c}{E$_{\mbox{\scriptsize{}tot}}$} &\multicolumn{1}{c}{E$_{\mbox{\scriptsize{}Ion}}$}  & \multicolumn{1}{c}{E$_{\mbox{\scriptsize{}Lit}}$}  &  \multicolumn{1}{c}{E$_{\mbox{\scriptsize{}tot}}$} &\multicolumn{1}{c}{E$_{\mbox{\scriptsize{}Ion}}$}    \\
\hline
 0.000  &-7.407126  &0.129935   &-7.41016\footnotemark[1]   & -7.334196  &0.057005   &     -7.33716\footnotemark[1]& -7.307804  &0.030612     &-7.31190 \footnotemark[1]   & -7.306793  &0.029602  \\
 0.001  &-7.408174  &0.130986   &-7.36609\footnotemark[2]   & -7.335244  &0.058055   &                             & -7.309562  &0.032373     &                            & -7.307796  &0.030607  \\
 0.010  &-7.416994  &0.139667   &-7.37481\footnotemark[2]   & -7.342662  &0.065336   &                             & -7.315155  &0.037828     &                            & -7.312592  &0.035266  \\
 0.050  &-7.451086  &0.173750   &                           & -7.356351  &0.079016   &                             & -7.329477  &0.052141     &                            & -7.309611  &0.032275  \\
 0.100  &-7.484773  &0.207876   &-7.4869343\footnotemark[3] & -7.360814  &0.083917   &                             & -7.328362  &0.051465     &                            & -7.310557  &0.033660  \\
 0.200  &-7.536032  &0.261359   &-7.49220\footnotemark[2]   & -7.367931  &0.093258   &                             & -7.322619  &0.047946     &                            & -7.302308  &0.027635  \\
 0.500  &-7.634547  &0.375024   &-7.6362483\footnotemark[3] & -7.372074  &0.112552   &                             & -7.312484  &0.052961     &                            & -7.285485  &0.025962  \\
 1.000  &-7.716679  &0.511132   &-7.66653\footnotemark[2]   & -7.335940  &0.130393   &                             & -7.263503  &0.057956     &                            & -7.236899  &0.031352  \\
 2.000  &-7.715709  &0.711256   &-7.66246\footnotemark[2]   & -7.154170  &0.149717   &                             & -7.067417  &0.062964     &                            & -7.038129  &0.033676  \\
 5.000  &-7.002346  &1.110399   &-6.94230\footnotemark[2]   & -6.068381  &0.176434   &                             & -5.961689  &0.069743     &                            & -5.927829  &0.035882  \\
 5.400  &-6.855410  &1.151263   &-6.8361629\footnotemark[3] & -5.882659  &0.178512   &                             & -5.774341  &0.070194     &                            & -5.740265  &0.036118  \\
10.00  &-4.684076  &1.530623   &-4.61777\footnotemark[2]   & -3.348774   &0.19532    &                             & -3.227581  &0.074128     &                            & -3.191029  &0.037576  \\
\end{tabular}
\caption{Total energies E$_{\mbox{\scriptsize Tot}}$, ionization
  energies E$_{\mbox{\scriptsize Ion}}$, and
previously published data E$_{\mbox{\scriptsize Lit}}$ in atomic units for the states
$\nu^2(-1)^+$ ($\nu=1,2,3,4$) at different field strengths $\gamma$.}\label{tab:n2-1+}  \footnotetext[1]{Ref. \cite{Sims:1975_1}}  \footnotetext[2]{Ref. \cite{Ivanov:1998_1}}  \footnotetext[3]{Ref. \cite{Guan:2001_1}}
\end{ruledtabular}
\end{table*}

\subsection{The symmetry subspace $^4(-1)^+$}
\label{sec:n4-1+}

\begin{figure}[htbp]
  \centering
  \includegraphics*[scale=0.30]{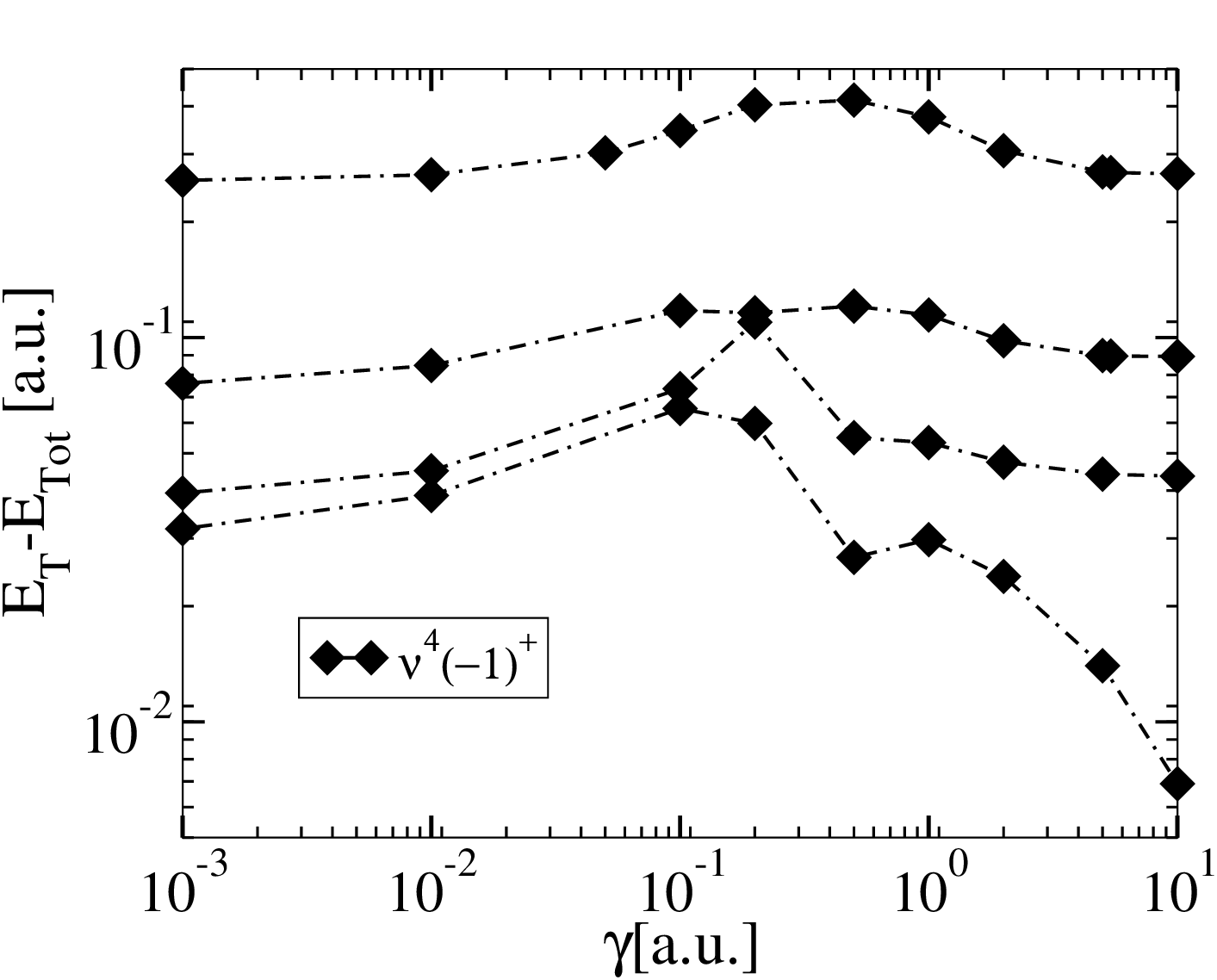}
  \caption{One-particle ionization energy for the states $\nu^4(-1)^+$ ($\nu=1,2,3,4$) in a.u. as a function of the magnetic field strength $\gamma$.}
  \label{fig:li_m-1q+}
\end{figure}

Our results for the symmetry subspace $^4(-1)^+$ are
presented in Fig.~\ref{fig:li_m-1q+} and in table~\ref{tab:n4-1+}. The
spin Zeeman-term causes the total energies
of all these fully spin polarized quartet states, to decrease
monotonously. 
On the other hand, this is not reflected by the one-particle
ionization energies, which increase or decrease weakly. The ground
state $1^4(-1)^+$ is a doubly tightly  
bound state predominantly consisting of the configuration $1s2s2p_{-1}$. An increase of the one-particle ionization energy can be
observed in the field regime $\gamma<0.2$, which is similar to the
increase in the ionization energy for the triply tightly bound state
$1^2(-1)^+$. However, in the field regime
$\gamma\ge0.5$ the ionization energy of the state $1^4(-1)^+$ decreases. The
reason for this are different one-particle ionization thresholds in
the different field regimes.
 For weak field strengths the ionization threshold involves  the Li$^+$ state $1^30^+$, which is a singly tightly
bound state,
 whereas for $\gamma>0.2$ it
is the $1^3(-1)^+$ state of the Li$^+$ ion which 
is a doubly tightly bound state. 

\begin{table*}[htpb]
\begin{ruledtabular}
\begin{tabular}{c*{3}{c}*{9}{c}}
& &\multicolumn{3}{c}{\rule[-0.7em]{0mm}{2.0em}{}$1^4(-1)^+$}& \multicolumn{2}{c}{$2 ^4(-1)^+$}& \multicolumn{2}{c}{$3 ^4(-1)^+$}& \multicolumn{2}{c}{$4 ^4(-1)^+$}\\
\hline
\multicolumn{1}{c}{\rule[-0.7em]{0mm}{2.0em}{}$\gamma $}& \multicolumn{1}{c}{T$_{\mbox{\scriptsize{}Sym}}$} &\multicolumn{1}{c}{E$_{\mbox{\scriptsize{}tot}}$} &\multicolumn{1}{c}{E$_{\mbox{\scriptsize{}Ion}}$}  & \multicolumn{1}{c}{E$_{\mbox{\scriptsize{}Lit}}$}   & \multicolumn{1}{c}{E$_{\mbox{\scriptsize{}tot}}$} &\multicolumn{1}{c}{E$_{\mbox{\scriptsize{}Ion}}$}  &  \multicolumn{1}{c}{E$_{\mbox{\scriptsize{}tot}}$} &\multicolumn{1}{c}{E$_{\mbox{\scriptsize{}Ion}}$}  & \multicolumn{1}{c}{E$_{\mbox{\scriptsize{}tot}}$} &\multicolumn{1}{c}{E$_{\mbox{\scriptsize{}Ion}}$}    \\
\hline
 0.000  & $     ^30^+$ & -5.366705    & 0.256072 & -5.35888\footnotemark[1]  &  -5.185835  & 0.075202    &   -5.149221  & 0.038588    & -5.141528  & 0.030895    \\
 0.001  & $     ^30^+$ & -5.368015    & 0.256375 & -5.36088\footnotemark[1]  &  -5.187601  & 0.075961    &   -5.151053  & 0.039413    & -5.143423  & 0.031783    \\
 0.010  & $     ^30^+$ & -5.385841    & 0.265227 & -5.37871\footnotemark[1]  &  -5.205247  & 0.084633    &   -5.165602  & 0.044987    & -5.159412  & 0.038798    \\
 0.100  & $     ^30^+$ & -5.550268    & 0.345788 & -5.54149\footnotemark[1]  &  -5.321977  & 0.117497    &   -5.278074  & 0.073594    & -5.269906  & 0.065426    \\
 0.200  & $  ^3(-1)^+$ & -5.703511    & 0.403056 & -5.69451\footnotemark[1]  &  -5.416500  & 0.116045    &   -5.410188  & 0.109732    & -5.360211  & 0.059756    \\
 0.500  & $  ^3(-1)^+$ & -6.058463    & 0.414736 & -6.04787\footnotemark[1]  &  -5.764438  & 0.120712    &   -5.698644  & 0.054917    & -5.670565  & 0.026839    \\
 1.000  & $  ^3(-1)^+$ & -6.494196    & 0.374980 & -6.48029\footnotemark[1]  &  -6.233729  & 0.114513    &   -6.172480  & 0.053263    & -6.149065  & 0.029849    \\
 2.000  & $  ^3(-1)^+$ & -7.206026    & 0.306258 & -7.18889\footnotemark[1]  &  -6.997908  & 0.098140    &   -6.947211  & 0.047444    & -6.923717  & 0.023949    \\
 5.000  & $  ^3(-1)^+$ & -8.905985    & 0.269712 & -8.88981\footnotemark[1]  &  -8.726111  & 0.089838    &   -8.680340  & 0.044067    & -8.650224  & 0.013952    \\
 5.400  & $  ^3(-1)^+$ & -9.096395    & 0.268724 & -9.0035\footnotemark[2]   & -8.917195   & 0.089525    &   -8.861757  & 0.034086    &   &   &  \\
10.00  & $  ^3(-1)^+$ & -10.925976   & 0.266916 & -10.91059\footnotemark[1] & -10.748366  & 0.0893068    &   -10.702625 & 0.043565    & -10.665919 & 0.006859    \\
\end{tabular}
\caption{Total energies E$_{\mbox{\scriptsize Tot}}$,  one-particle
  ionization energies E$_{\mbox{\scriptsize Ion}}$, and
previously published data E$_{\mbox{\scriptsize Lit}}$ in a. u. for the states
$\nu^4(-1)^+$ ($\nu=1,2,3,4$), as well as the threshold symmetry T$_{\mbox{\scriptsize{}Sym}}$ for different field strengths
$\gamma$.}\label{tab:n4-1+}  \footnotetext[1]{Ref. \cite{Ivanov:1998_1}.}  \footnotetext[2]{Ref. \cite{Jones:1996_1}.}  
\end{ruledtabular}
\end{table*}

The decrease in the one-particle ionization threshold can be also
observed for the higher excited states of this symmetry
($\nu^4(-1)^+$ with $\nu=2,3,4$). Additional avoided crossings cause
the one-particle 
ionization energies of the states $3^4(-1)^+$ and $4^4(-1)^+$ to
decrease. This is very impressive for the state $4^4(-1)^+$, for which the
ionization energy at $\gamma=0.1$  is $0.065426$~a.u. and decreases about one
order of magnitude   to $0.006859$~a.u. at $\gamma=10$.

\subsection{The symmetry subspace $\nu^4(-1)^-$}
\label{sec:n4-1-}
\begin{figure}[htbp]
  \centering
  \includegraphics*[scale=0.30]{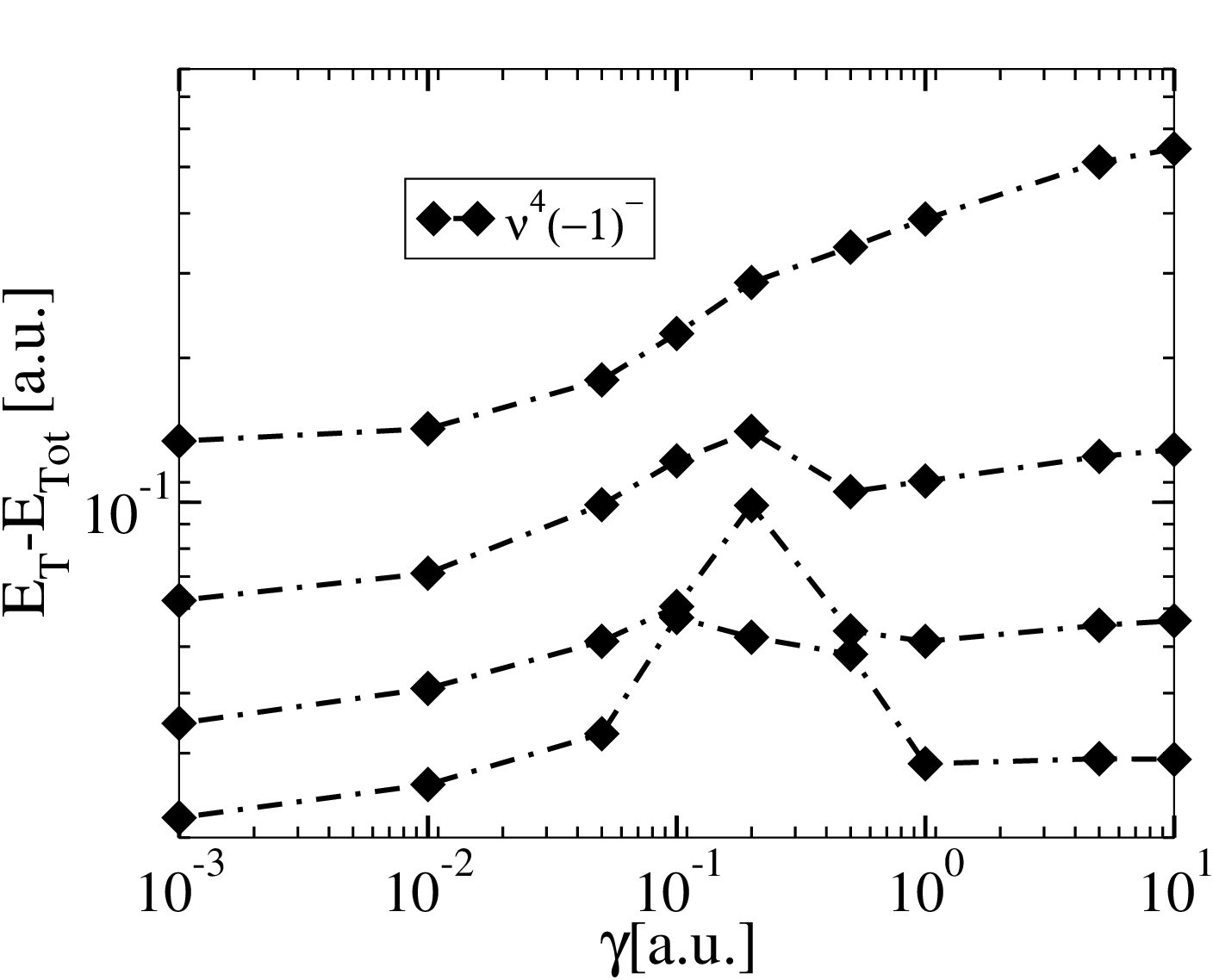}
  \caption{One-particle ionization energy for the states $\nu^4(-1)^-$ ($\nu=1,2,3,4$) in a.u. as a function of the magnetic field strength $\gamma$.}
  \label{fig:li_m-1q-}
\end{figure}

In this subsection  we will review the results for the quartet states
with magnetic 
quantum number $M=-1$ and negative $z$-parity, i.e. $\nu^4(-1)^-$
($\nu=1,2,3,4$).  In the low field regime ($\gamma<0.1$) all the
curves in Fig. \ref{fig:li_m-1q-} behave similary: we observe a
significant increase of the ionization energies for all considered
states of this symmetry subspace. At higher field
strengths the energies develop differently for the different states. The
one-particle ionization energy of the state $1^4(-1)^-$  increases
monotonously. The ionization energies of the higher excited states $2^4(-1)^-$,
$3^4(-1)^-$ and $4^4(-1)^-$ reach a local maximum for
$0.1<\gamma<1$. At higher field strengths ($\gamma>1$) we
observe, that the one-particle
ionization energies for these states become nearly field independent.

\begin{table*}[htpb]
\begin{ruledtabular}
\begin{tabular}{c*{3}{c}*{9}{c}}
& &\multicolumn{3}{c}{\rule[-0.7em]{0mm}{2.0em}{}$1^4(-1)^-$}& \multicolumn{2}{c}{$2 ^4(-1)^-$}& \multicolumn{2}{c}{$3 ^4(-1)^-$}& \multicolumn{2}{c}{$4 ^4(-1)^-$}\\
\hline
\multicolumn{1}{c}{\rule[-0.7em]{0mm}{2.0em}{}$\gamma $}& \multicolumn{1}{c}{T$_{\mbox{\scriptsize{}Sym}}$} &\multicolumn{1}{c}{E$_{\mbox{\scriptsize{}tot}}$} &\multicolumn{1}{c}{E$_{\mbox{\scriptsize{}Ion}}$}  & \multicolumn{1}{c}{E$_{\mbox{\scriptsize{}Lit}}$}   & \multicolumn{1}{c}{E$_{\mbox{\scriptsize{}tot}}$} &\multicolumn{1}{c}{E$_{\mbox{\scriptsize{}Ion}}$}  &  \multicolumn{1}{c}{E$_{\mbox{\scriptsize{}tot}}$} &\multicolumn{1}{c}{E$_{\mbox{\scriptsize{}Ion}}$}  & \multicolumn{1}{c}{E$_{\mbox{\scriptsize{}tot}}$} &\multicolumn{1}{c}{E$_{\mbox{\scriptsize{}Ion}}$}    \\
\hline
 0.000  & $     ^30^+$ &-5.243519    &0.1328852   &-5.24554\footnotemark[1]  & -5.172069  & 0.0614360    &  -5.144236    & 0.0336028    & -5.128015  &  0.0173817   \\
 0.001  & $     ^30^+$ &-5.245744    &0.1341042   &-5.23386\footnotemark[2]  & -5.174078  & 0.0624380    &  -5.146267    & 0.0346276    & -5.133621  &  0.0219813   \\
 0.010  & $     ^30^+$ &-5.262918    &0.1423040   &-5.25170\footnotemark[2]  & -5.191736  & 0.0711218    &  -5.161532    & 0.0409179    & -5.146381  &  0.0257669   \\
 0.050  & $     ^30^+$ &-5.339046    &0.1799396   &                          & -5.257953  & 0.0988460    &  -5.210369    & 0.0512629    & -5.192019  &  0.0329123   \\
 0.100  & $     ^30^+$ &-5.429067    &0.2245868   &-5.41643\footnotemark[2]  & -5.326415  & 0.1219344    &  -5.265077    & 0.0605963    & -5.262003  &  0.0575226   \\
 0.200  & $  ^3(-1)^+$ &-5.587442    &0.2869864   &-5.57585\footnotemark[2]  & -5.441002  & 0.1405464    &  -5.399005    & 0.0985497    & -5.352709  &  0.0522539   \\
 0.500  & $  ^3(-1)^+$ &-5.983849    &0.3401224   &-5.96957\footnotemark[2]  & -5.748951  & 0.1052242    &  -5.697593    & 0.0538666    & -5.691942  &  0.0482154   \\
 1.000  & $  ^3(-1)^+$ &-6.508527    &0.3893105   &-6.49248\footnotemark[2]  & -6.229901  & 0.1106850    &  -6.170544    & 0.0513275    & -6.147671  &  0.0284546   \\
 5.000  & $  ^3(-1)^+$ &-9.148122    &0.5118489   &-9.12554\footnotemark[2]  & -8.760873  & 0.1245998    &  -8.691638    & 0.0553655    & -8.665490  &  0.0292174   \\
10.00  & $  ^3(-1)^+$ &-11.204311   &0.5452518   &-11.17886\footnotemark[2] & -10.787789 & 0.1287291    &  -10.715697   & 0.0566372    & -10.688192 &  0.0291328   \\
\end{tabular}
\caption{Total energies E$_{\mbox{\scriptsize Tot}}$,  one-particle ionization
  energies E$_{\mbox{\scriptsize Ion}}$, and
previously published data E$_{\mbox{\scriptsize Lit}}$ in a. u. for the states
$\nu^4(-1)^-$ ($\nu=1,2,3,4$) and the corresponding threshold
  symmetry T$_{\mbox{\scriptsize{}Sym}}$ at different field strengths $\gamma$.}\label{tab:n4-1-}  \footnotetext[1]{Ref. \cite{Ivanov:1998_1}.}  \footnotetext[2]{Ref. \cite{Jones:1996_1}.}  
\end{ruledtabular}
\end{table*}

Table~\ref{tab:n4-1-} contains the corresponding numerical values. For
this symmetry  
subspace  a crossover for the Li$^+$ threshold state can be observed, as
for the $^4(-1)^+$ subspace, discussed in the previous subsection.

\subsection{The symmetry subspace $\nu^2(-2)^+$}
\label{sec:n2-2+}

\begin{figure}[htbp]
  \centering
  \includegraphics*[scale=0.30]{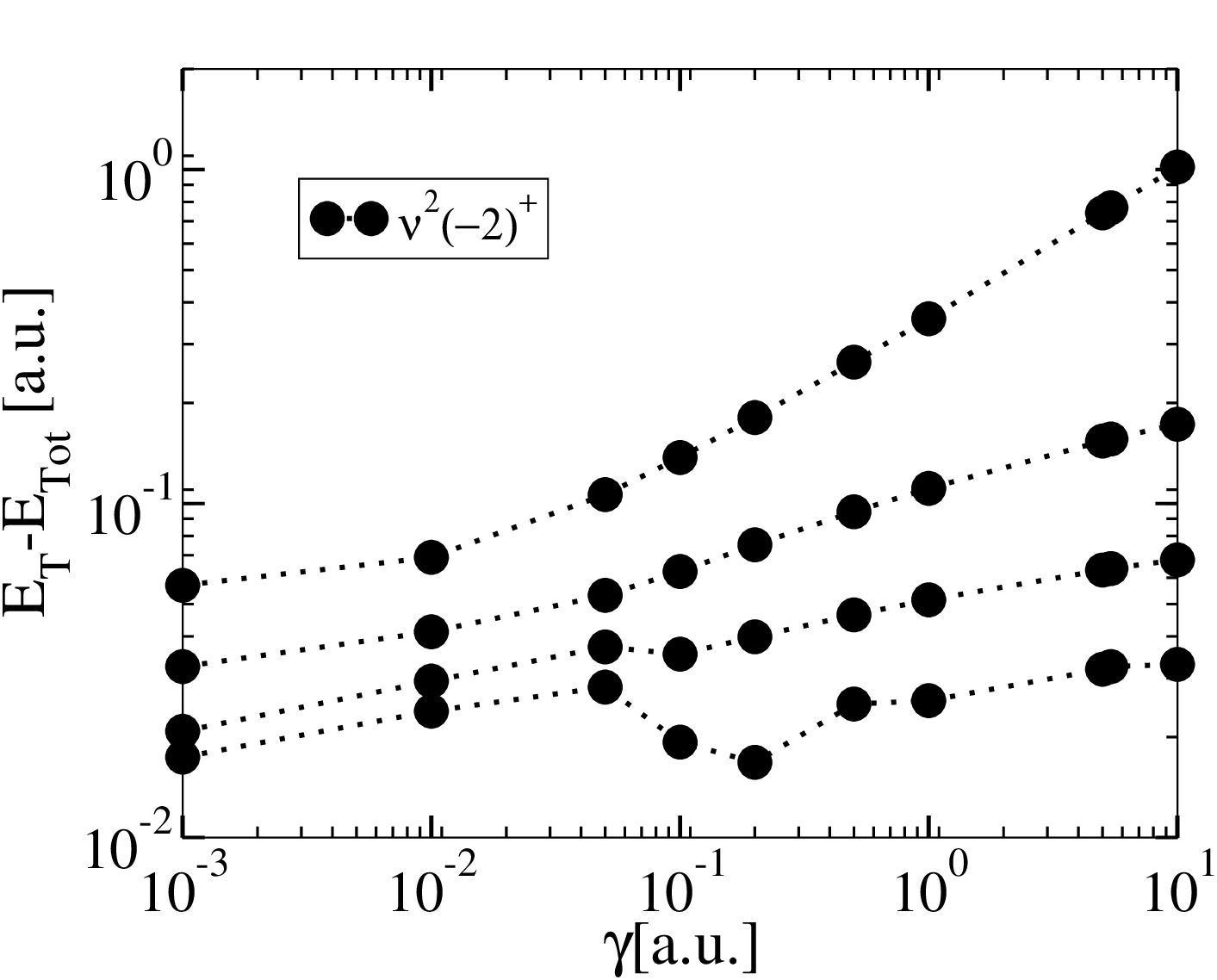}
  \caption{One-particle ionization energy for the states $\nu^2(-2)^+$ ($\nu=1,2,3,4$) in a.u. as a function of the magnetic field strength $\gamma$.}
  \label{fig:li_m-2d+}
\end{figure}

In this subsection, we present our results for the
$^2(-2)^+$ symmetry subspace. Figure~\ref{fig:li_m-2d+}
 shows the curves for the one-particle ionization energy and 
table~\ref{tab:n2-2+} contains  the numerical results for the
total energies, one-particle ionization energies as well as previously
published data.  The ground state of this symmetry $1^2(-2)^+$ is a
triply
tightly bound state. Therefore the
one-particle ionization energy increases monotonously. For this state
the increase amounts to more than one order of
magnitude in the considered range of field strengths. At zero field it is
$0.0554$~a.u. and at $\gamma=10$ $1.017$~a.u., i.e.
an  increase by approximately a factor of 20.
\begin{table*}[htpb]
\begin{ruledtabular}
\begin{tabular}{c*{2}{c}*{7}{c}}
 &\multicolumn{3}{c}{\rule[-0.7em]{0mm}{2.0em}{}$1^2(-2)^+$}& \multicolumn{2}{c}{$2 ^2(-2)^+$}& \multicolumn{2}{c}{$3 ^2(-2)^+$}& \multicolumn{2}{c}{$4 ^2(-2)^+$}\\
\hline
\multicolumn{1}{c}{\rule[-0.7em]{0mm}{2.0em}{}$\gamma $} &\multicolumn{1}{c}{E$_{\mbox{\scriptsize{}tot}}$} &\multicolumn{1}{c}{E$_{\mbox{\scriptsize{}Ion}}$}  & \multicolumn{1}{c}{E$_{\mbox{\scriptsize{}Lit}}$}   & \multicolumn{1}{c}{E$_{\mbox{\scriptsize{}tot}}$} &\multicolumn{1}{c}{E$_{\mbox{\scriptsize{}Ion}}$}  &  \multicolumn{1}{c}{E$_{\mbox{\scriptsize{}tot}}$} &\multicolumn{1}{c}{E$_{\mbox{\scriptsize{}Ion}}$}  & \multicolumn{1}{c}{E$_{\mbox{\scriptsize{}tot}}$} &\multicolumn{1}{c}{E$_{\mbox{\scriptsize{}Ion}}$}    \\
\hline
 0.000  & -7.332617  &0.055426   &-7.335523541\footnotemark[1] & -7.308297 & 0.031106    &  -7.297036   & 0.019845    & -7.296823  & 0.019632     \\
 0.001  & -7.334097  &0.056908   &                             & -7.309735 & 0.032547    &  -7.297990   & 0.020801    & -7.294566  & 0.017378     \\
 0.010  & -7.346296  &0.068969   &                             & -7.318595 & 0.041269    &  -7.306747   & 0.029421    & -7.301231  & 0.023905     \\
 0.050  & -7.383648  &0.106312   &                             & -7.330479 & 0.053143    &  -7.314511   & 0.037175    & -7.305505  & 0.028169     \\
 0.100  & -7.414207  &0.137310   &-7.4169780 \footnotemark[2]  & -7.339471 & 0.062574    &  -7.312327   & 0.035430    & -7.296226  & 0.019329     \\
 0.200  & -7.455585  &0.180912   &                             & -7.349825 & 0.075152    &  -7.314596   & 0.039923    & -7.291517  & 0.016844     \\
 0.500  & -7.524481  &0.264958   &                             & -7.353984 & 0.094462    &  -7.305925   & 0.046403    & -7.284644  & 0.025122     \\
 1.000  & -7.562892  &0.357345   &                             & -7.316557 & 0.111010    &  -7.257044   & 0.051497    & -7.231281  & 0.025735     \\
 5.000  & -6.633118  &0.741172   &                             & -6.045774 & 0.153827    &  -5.955127   & 0.063181    &  -5.923863 & 0.031917    \\
 5.400  & -6.472203  &0.768057   &-6.451608 \footnotemark[2]   & -5.860240 & 0.156093    &  -5.768012   & 0.063865    &  -5.736691 & 0.032544    \\
10.00  & -4.170890  &1.017437   &                             & -3.326137  & 0.172684    &  -3.221213   & 0.067760    &  -3.186407 & 0.032955    \\
\end{tabular}
\caption{Total energies E$_{\mbox{\scriptsize Tot}}$,  one-particle
  ionization energies E$_{\mbox{\scriptsize Ion}}$, and
previously published data E$_{\mbox{\scriptsize Lit}}$ in atomic units for the states
$\nu^2(-2)^+$ ($\nu=1,2,3,4$) at different field strengths $\gamma$.}\label{tab:n2-2+}  \footnotetext[1]{Ref. \cite{Yan:1995_1}.}  \footnotetext[2]{Ref.\cite{Guan:2001_1} .}  
\end{ruledtabular}
\end{table*}

For the first excited state of this symmetry ($2^2(-2)^+$) we observe
in Fig.~\ref{fig:li_m-2d+} an increase, which is less pronounced than for
the ground state of the same symmetry. We obtain for  its one-particle ionization energy
 at vanishing field $0.031106$~a.u. and at the highest considered field
strength ($\gamma=10$)  $0.172684$~a.u., which corresponds to an
increase of a factor 5.

For the two higher excited states of this symmetry ($3^2(-2)^+$ and
$4^2(-2)^+$) the situation is different. In the intermediate field
regime again avoided crossings take place. Therefore their one-particle
ionization energies pass through local minima  at $\gamma
\approx0.1$ ($3^2(-2)^+$) and $\gamma\approx0.2$ ($4^2(-2)^+$) respectively.

\subsection{The symmetry subspace $\nu^4(-2)^+$}
\label{sec:n4-2+}

\begin{figure}[htbp]
  \centering
  \includegraphics*[scale=0.30]{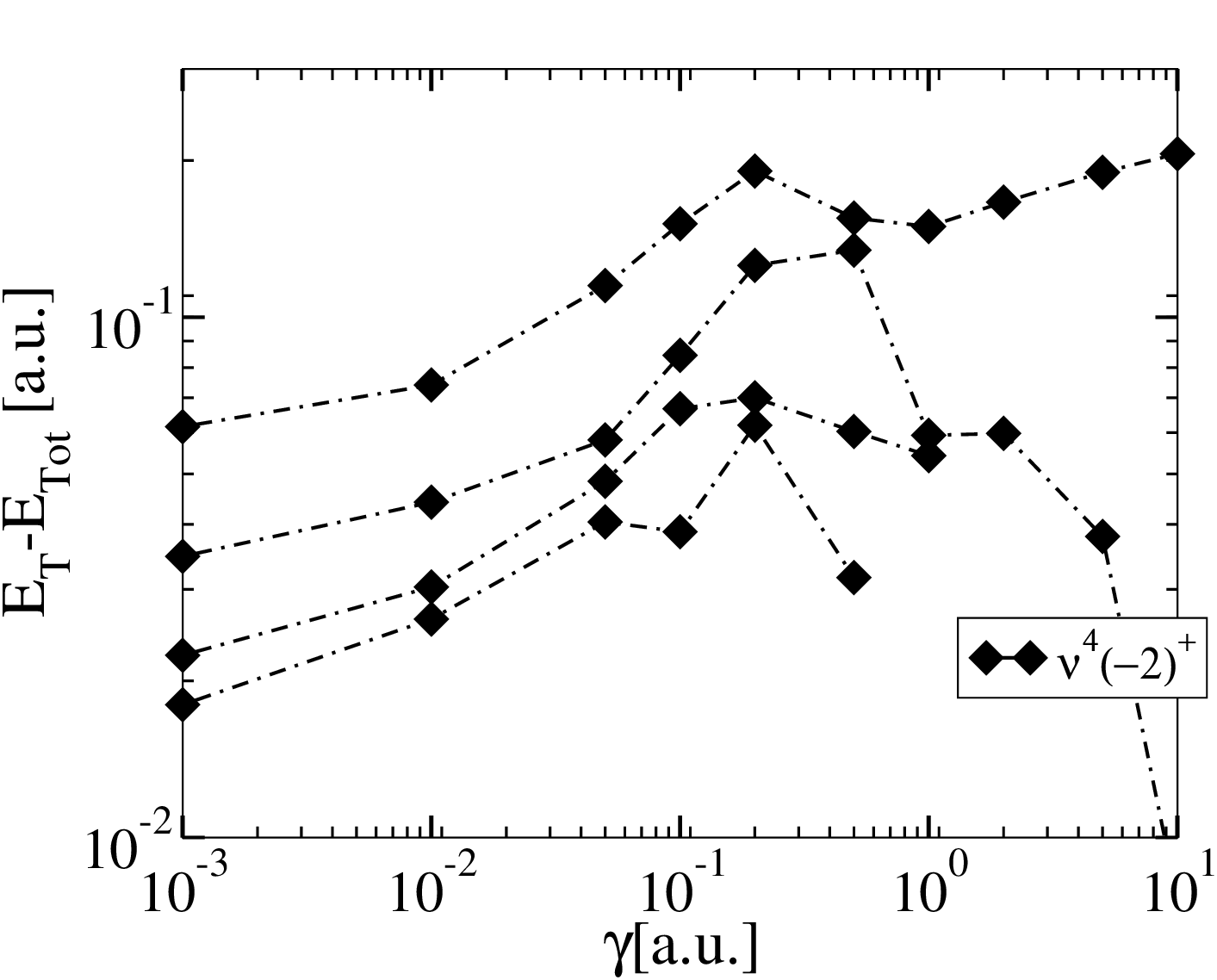}
  \caption{One-particle ionization energies for the states $\nu^4(-2)^+$ ($\nu=1,2,3,4$) in a.u. as a function of the magnetic field strength $\gamma$.}
  \label{fig:li_m-2q+}
\end{figure}

For the corresponding quadruplet subspace $^4(-2)^+$ the ionization
energies  are shown  in Fig.~\ref{fig:li_m-2q+} and numerical values
are given in
table~\ref{tab:n4-2+}. The  behavior of the
one-particle ionization energies is different compared  to the
corresponding behavior of the  doublet 
states presented in the previous subsection. At low fields
($\gamma<0.05$) an increase  can be observed for all
states considered in this work. For the ground state $1^4(-2)^+$ the
one-particle ionization energy increases from $0.06017$~a.u. at
$\gamma=0$ to $0.190927$~a.u.  at $\gamma=0.2$, where it reaches a
local maximum. At this field strength the ionization threshold
changes, as for the other quadruplet states. Consequently the ionization
energy decreases and passes through a local minimum at
$\gamma\approx1$. At $\gamma\approx0.5$ an avoided crossing occurs, which leads to an increase of the one-particle ionization
energy for the state $1^4(-2)^+$ for higher field strengths. On the
other hand, the corresponding energy for the state $2^4(-2)^+$ which
increases for $\gamma\gtrsim0.5$ acquires a strongly decreasing
behavior due to this avoided crossing. Further avoided crossings
among the higher excited states cause the  states
$3^4(-2)^+$ and $4^4(-2)^+$ to become unbound for $\gamma>1$
($3^4(-2)^+$) and $\gamma>0.5$ ($4^4(-2)^+$), respectively. We remark,
that there are no prevoiusly calculated data on states of the
$^4(-2)^+$ symmetry in the literature.

\begin{table*}[htpb]
\begin{ruledtabular}
\begin{tabular}{*{10}{c}}
& &\multicolumn{2}{c}{\rule[-0.7em]{0mm}{2.0em}{}$1^4(-2)^+$}& \multicolumn{2}{c}{$2 ^4(-2)^+$}& \multicolumn{2}{c}{$3 ^4(-2)^+$}& \multicolumn{2}{c}{$4 ^4(-2)^+$}\\
\hline
\multicolumn{1}{c}{\rule[-0.7em]{0mm}{2.0em}{}$\gamma $}& \multicolumn{1}{c}{T$_{\mbox{\scriptsize{}Sym}}$} &\multicolumn{1}{c}{E$_{\mbox{\scriptsize{}tot}}$} &\multicolumn{1}{c}{E$_{\mbox{\scriptsize{}Ion}}$}   & \multicolumn{1}{c}{E$_{\mbox{\scriptsize{}tot}}$} &\multicolumn{1}{c}{E$_{\mbox{\scriptsize{}Ion}}$}  &  \multicolumn{1}{c}{E$_{\mbox{\scriptsize{}tot}}$} &\multicolumn{1}{c}{E$_{\mbox{\scriptsize{}Ion}}$}  & \multicolumn{1}{c}{E$_{\mbox{\scriptsize{}tot}}$} &\multicolumn{1}{c}{E$_{\mbox{\scriptsize{}Ion}}$}    \\
\hline
 0.000  & $     ^30^+$ & -5.170803   & 0.060170  & -5.143875  & 0.033242    & -5.131679    & 0.021046    & -5.124827 & 0.014194    \\
 0.001  & $     ^30^+$ & -5.173261   & 0.061621  & -5.146306  & 0.034666    & -5.134036    & 0.022396    & -5.129671 & 0.018031    \\
 0.010  & $     ^30^+$ & -5.194678   & 0.074063  & -5.164667  & 0.044053    & -5.150889    & 0.030275    & -5.146865 & 0.026251    \\
 0.050  & $     ^30^+$ & -5.274146   & 0.115039  & -5.217208  & 0.058102    & -5.207536    & 0.048430    & -5.199501 & 0.040395    \\
 0.100  & $     ^30^+$ & -5.355652   & 0.151172  & -5.289006  & 0.084526    & -5.271140    & 0.066660    & -5.243141 & 0.038661    \\
 0.200  & $  ^3(-1)^+$ & -5.491382   & 0.190927  & -5.426277  & 0.125822    & -5.370458    & 0.070003    & -5.362501 & 0.062046    \\
 0.500  & $  ^3(-1)^+$ & -5.798837   & 0.155110  & -5.778279  & 0.134552    & -5.704042    & 0.060316    & -5.675295 & 0.031569    \\
 1.000  & $  ^3(-1)^+$ & -6.268653   & 0.149437  & -6.178554  & 0.059338    & -6.173405    & 0.054189    &  &     \\
 2.000  & $  ^3(-1)^+$ & -7.066163   & 0.166395  & -6.959543  & 0.059775    &     &     &  &     \\
 5.000  & $  ^3(-1)^+$ & -8.826185   & 0.189912  & -8.674211  & 0.037938    &     &     &  &     \\
 10.00  & $  ^3(-1)^+$ & -10.865122  & 0.206062  & -10.666670 & 0.007610    &     &     &  &       \\
\end{tabular}
\caption{Total energies E$_{\mbox{\scriptsize Tot}}$, and one-particle ionization
  energies E$_{\mbox{\scriptsize Ion}}$ in a.u. for the states
$\nu^4(-2)^+$ ($\nu=1,2,3,4$) and the  threshold
  symmetry T$_{\mbox{\scriptsize{}Sym}}$ at different field strengths $\gamma$.}\label{tab:n4-2+}  
\end{ruledtabular}
\end{table*}

\subsection{The symmetry subspace $\nu^4(-3)^+$}
\label{sec:n4-3+}

\begin{figure}[htbp]
  \centering
  \includegraphics*[scale=0.30]{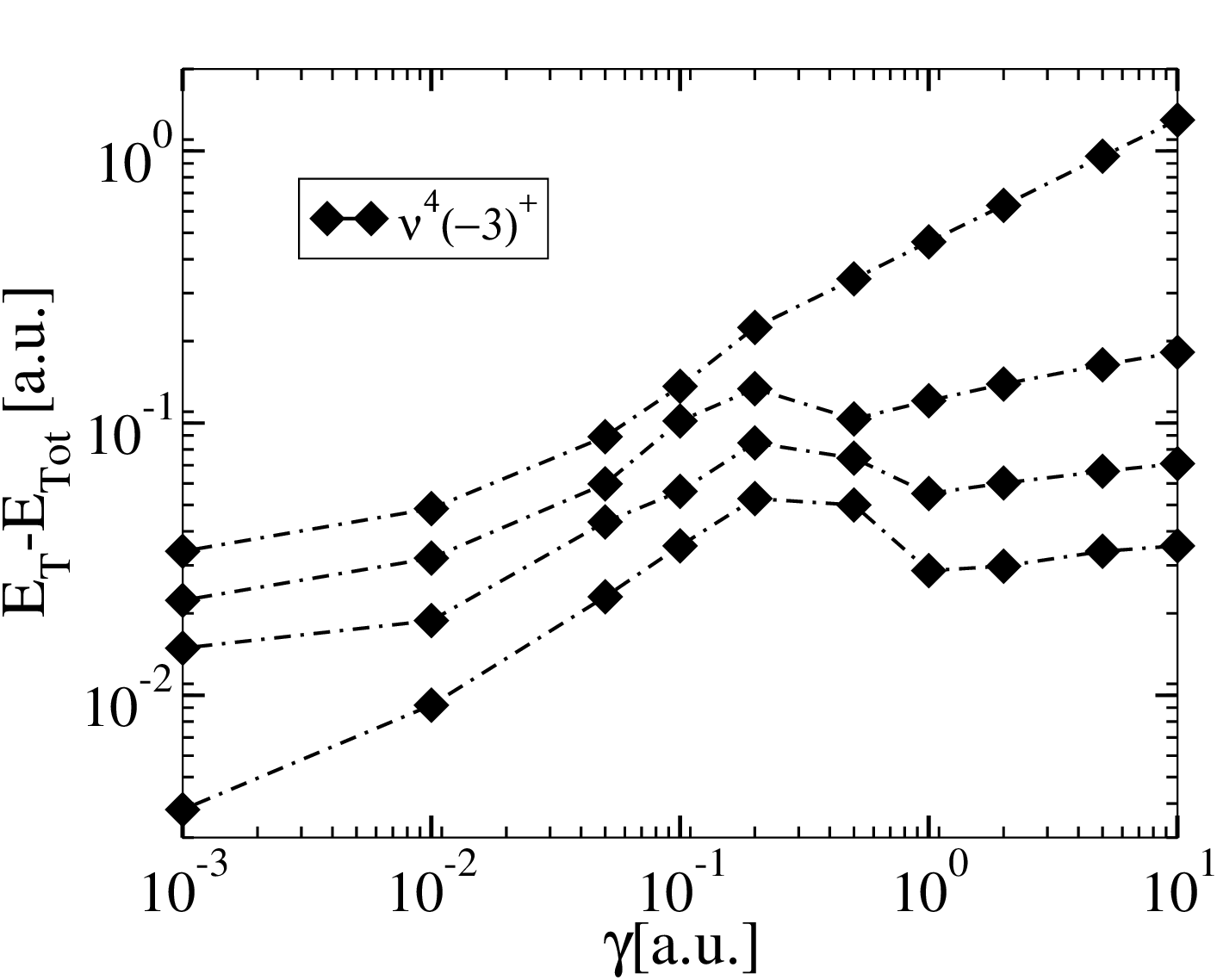}
  \caption{One-particle ionization energy for the states $\nu^4(-3)^+$ ($\nu=1,2,3,4$) in a.u. as a function of the magnetic field strength $\gamma$.}
  \label{fig:li_m-3q+}
\end{figure}

Let us discuss our  results for the symmetry
subspace $^{2S+1}M^{\Pi_z}=^4(-3)^+$. The energetically lowest state in this
subspace represents the global ground state of the lithium atom in the
high field regime, as mentioned above. It is a triply tightly bound
state containing a.o. the orbitals $1s2p_{-1}3d_{-2}$. In
Fig.~\ref{fig:li_m-3q+} it can be seen, that its one-particle 
ionization energy increases strongly as a function of the 
field strength. At zero magnetic field it is about $0.03164$~a.u., whereas at
$\gamma=10$ an energy of $1.3033$~a.u.  is needed to ionize
the state. Therefore the one-particle ionization energy increases
roughly by a factor of 40. However, if the reader compares the
one-particle ionization energies in
table~\ref{tab:n4-3+} with table~\ref{tab:n2-1+}, it is evident that the
state $1^4(-3)^+$ is not the  state with the highest ionization energy
for any field
strength. 
\begin{table*}[htpb]
\begin{ruledtabular}
\begin{tabular}{*{11}{c}}
& &\multicolumn{3}{c}{\rule[-0.7em]{0mm}{2.0em}{}$1^4(-3)^+$}& \multicolumn{2}{c}{$2 ^4(-3)^+$}& \multicolumn{2}{c}{$3 ^4(-3)^+$}& \multicolumn{2}{c}{$4 ^4(-3)^+$}\\
\hline
\multicolumn{1}{c}{\rule[-0.7em]{0mm}{2.0em}{}$\gamma $}& \multicolumn{1}{c}{T$_{\mbox{\scriptsize{}Sym}}$} &\multicolumn{1}{c}{E$_{\mbox{\scriptsize{}tot}}$} &\multicolumn{1}{c}{E$_{\mbox{\scriptsize{}Ion}}$}  &\multicolumn{1}{c}{E$_{\mbox{\scriptsize{}Lit}}$}  & \multicolumn{1}{c}{E$_{\mbox{\scriptsize{}tot}}$} &\multicolumn{1}{c}{E$_{\mbox{\scriptsize{}Ion}}$}  &  \multicolumn{1}{c}{E$_{\mbox{\scriptsize{}tot}}$} &\multicolumn{1}{c}{E$_{\mbox{\scriptsize{}Ion}}$}  & \multicolumn{1}{c}{E$_{\mbox{\scriptsize{}tot}}$} &\multicolumn{1}{c}{E$_{\mbox{\scriptsize{}Ion}}$}    \\
\hline
 0.000  & $     ^30^+$ & -5.142319   & 0.031686  & -5.08379\footnotemark[1]  &   -5.125979    & 0.015346    &     &  &   &   \\
 0.001  & $     ^30^+$ & -5.145464   & 0.033824  & -5.08679\footnotemark[1]  &   -5.133945    & 0.022305    & -5.126540     & 0.014901 &  -5.115403   & 0.003764  \\
 0.010  & $     ^30^+$ & -5.169111   & 0.048497  & -5.11268\footnotemark[1]  &   -5.152521    & 0.031907    & -5.139446     & 0.018831 &  -5.129767   & 0.009153  \\
 0.050  & $     ^30^+$ & -5.248206   & 0.089099  &                           &   -5.218924    & 0.059818    & -5.202450     & 0.043344 &  -5.182078   & 0.022972  \\
 0.100  & $     ^30^+$ & -5.341030   & 0.136549  & -5.32140\footnotemark[1]  &   -5.306295    & 0.101815    & -5.260565     & 0.056085 &  -5.239866   & 0.035386  \\
 0.200  & $  ^3(-1)^+$ & -5.524939   & 0.224483  & -5.51151\footnotemark[1]  &   -5.434168    & 0.133713    & -5.384936     & 0.084481 &  -5.353290   & 0.052835  \\
 0.500  & $  ^3(-1)^+$ & -5.982253   & 0.338527  & -5.97952\footnotemark[2]  &   -5.747212    & 0.103486    & -5.717997     & 0.074271 &  -5.693685   & 0.049959  \\
 1.000  & $  ^3(-1)^+$ & -6.582361   & 0.463144  & -6.57081\footnotemark[1]  &   -6.240001    & 0.120785    & -6.174336     & 0.055119 &  -6.147905   & 0.028689  \\
 2.000  & $  ^3(-1)^+$ & -7.530125   & 0.630357  & -7.52003\footnotemark[1]  &   -7.038917    & 0.139149    & -6.959997     & 0.060229 &  -6.929553   & 0.029786  \\
 5.000  & $  ^3(-1)^+$ & -9.591769   & 0.955496  & -9.57694\footnotemark[1]  &   -8.799910    & 0.163637    & -8.702726     & 0.066453 &  -8.670060   & 0.033787  \\
10.00  & $  ^3(-1)^+$ & -11.957294  & 1.298234  & -11.93902\footnotemark[1] &   -10.841017   & 0.181957    & -10.730105    & 0.071045 &  -10.694481  & 0.035421  \\
\end{tabular}
\caption{Total energies E$_{\mbox{\scriptsize Tot}}$,  one-particle ionization energies
  E$_{\mbox{\scriptsize Ion}}$, and
  previously published data E$_{\mbox{\scriptsize Lit}}$ in a.u., as
  well as , threshold symmetry
  T$_{\mbox{\scriptsize{}Sym}}$ for the states
$\nu^4(-3)^+$ ($\nu=1,2,3,4$) at different field strengths $\gamma$.}\label{tab:n4-3+}    \footnotetext[1]{Ref. \cite{Ivanov:1998_1}.}  \footnotetext[2]{Ref. \cite{Ivanov:2000_1}.} 
\end{ruledtabular}
\end{table*}

The ionization energies for the  excited states $2^4(-3)^+$, $3^4(-3)^+$
and $4^4(-3)^+$ behave all very similar. Their
ionization energy increases up to $\gamma\approx 0.5$, where a local
maximum is reached. For
$0.5\gtrsim\gamma\gtrsim2$ the ionization energy decreases,
whereas for higher field strengths it increases
again with a lower slope than for $\gamma<0.5$. Our numerical energy
values for the $1^4(-3)^+$ state are always lower than the energies
obtained in the literature (see table~\ref{tab:n4-3+} and
Refs.~\cite{Ivanov:1998_1,Ivanov:2000_1}).

\section{Wavelengths for  electromagnetic transitions}
\label{sec:wave}

\begin{figure}[htbp]
  \centering
  \includegraphics*[scale=0.3]{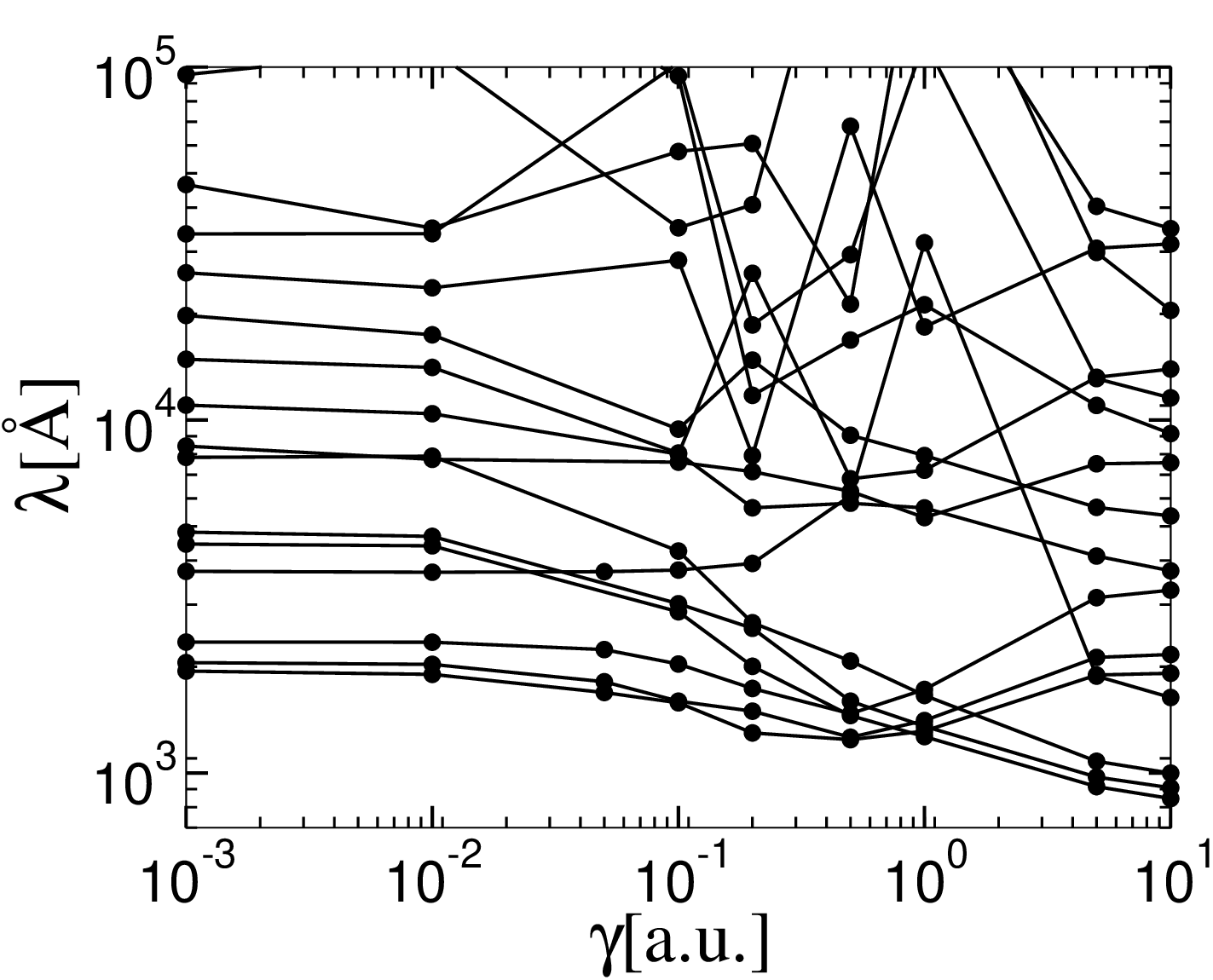}
  \caption{Transition wavelengths $\lambda$ for the linear polarized
  transitions $\nu^4(-1)^+\longrightarrow\mu^4(-1)^-$ ($\nu,\mu=1,2,3,4$)
  in \AA{} as a function of the magnetic field strength $\gamma$.}
  \label{fig:lt_trans_-1_0_-1_1}
\end{figure}

\begin{figure}[htbp]
  \centering
  \includegraphics*[scale=0.3]{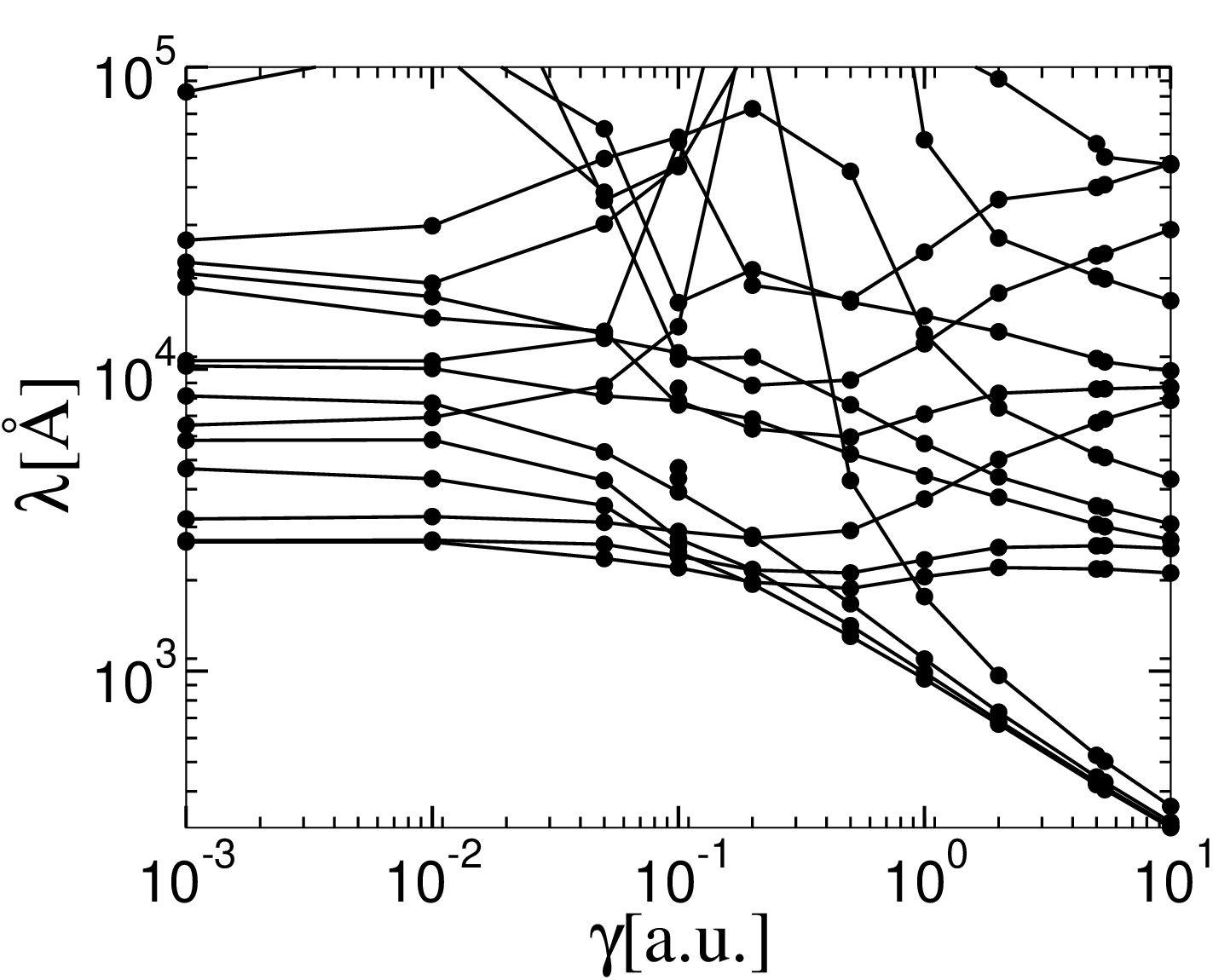}
  \caption{Transition wavelengths $\lambda$ for the circular polarized
  transitions $\nu^20^+\longrightarrow\mu^2(-1)^+$ ($\nu,\mu=1,2,3,4$)
  in \AA{} as a function of the magnetic field strength $\gamma$.}
  \label{fig:lt_trans_0_0_-1_0}
\end{figure}

\begin{figure}[htbp]
  \centering
  \includegraphics*[scale=0.3]{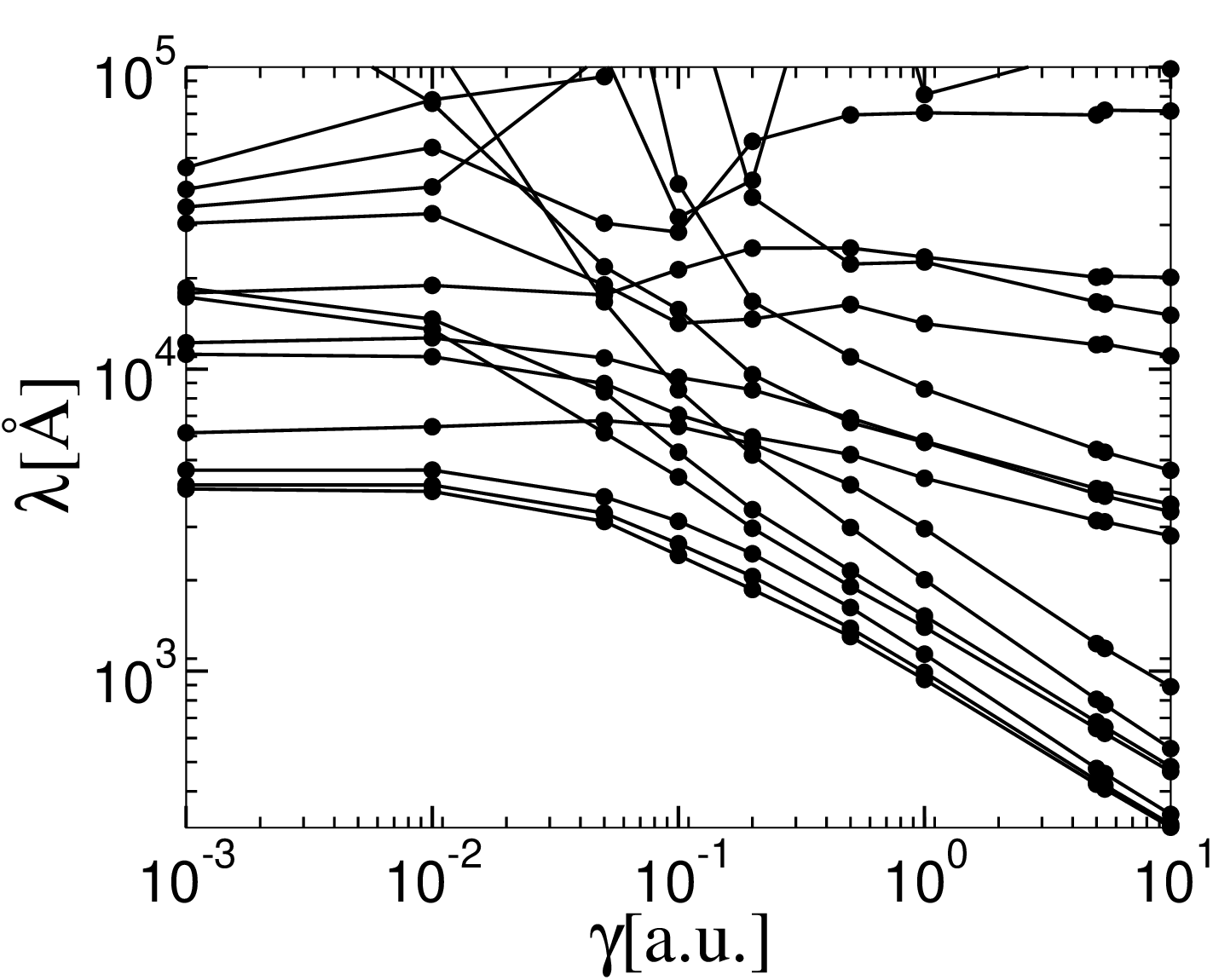}
  \caption{Transition wavelengths $\lambda$ for the circular polarized
  transitions $\nu^2(-1)^+\longrightarrow\mu^2(-2)^+$ ($\nu,\mu=1,2,3,4$)
  in \AA{} as a function of the magnetic field strength $\gamma$.}
  \label{fig:lt_trans_-1_0_-2_0_dubl}
\end{figure}

\begin{figure}[htbp]
  \centering
  \includegraphics*[scale=0.3]{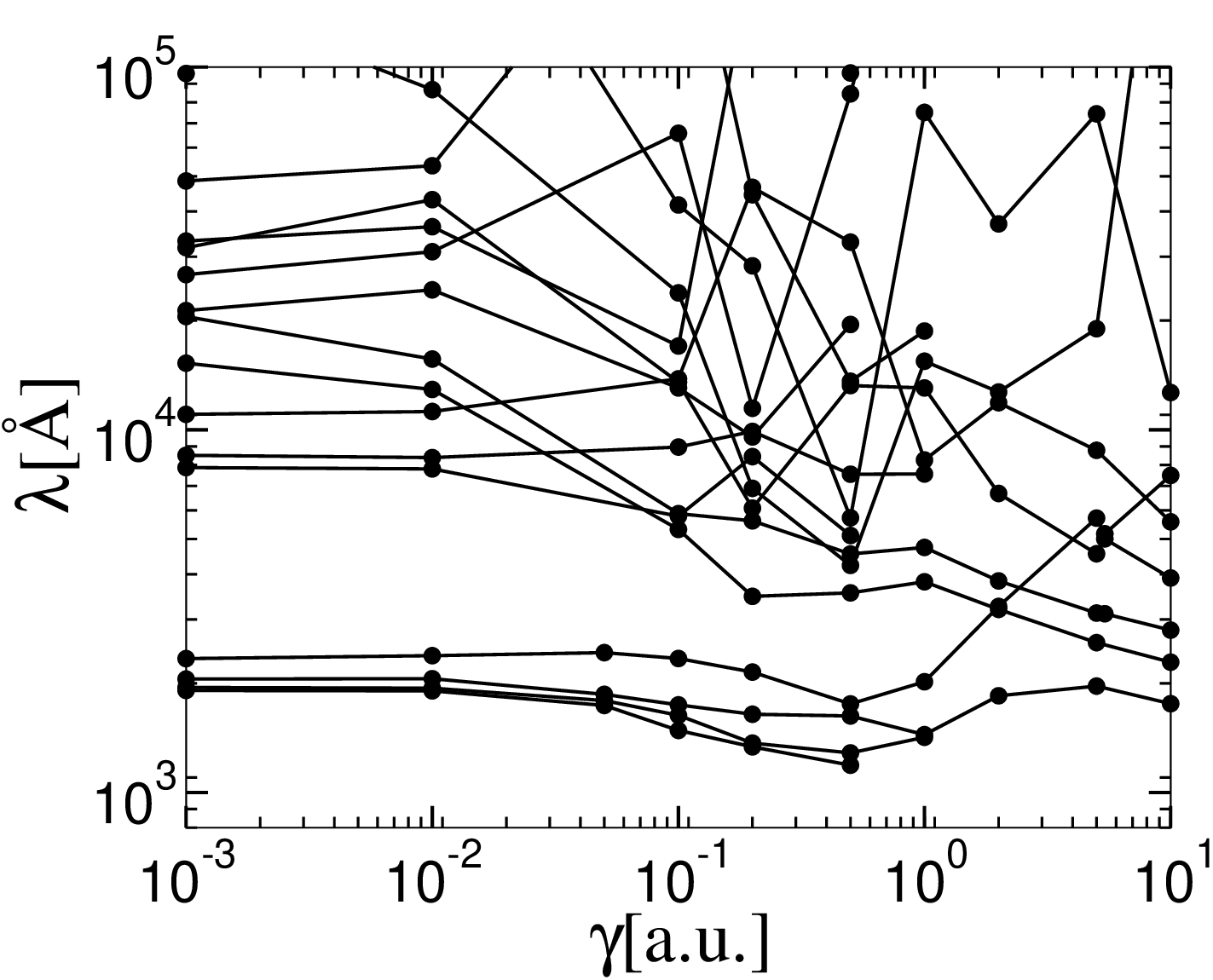}
  \caption{Transition wavelengths $\lambda$ for the circular polarized
  transitions $\nu^4(-1)^+\longrightarrow\mu^4(-2)^+$ ($\nu,\mu=1,2,3,4$)
  in \AA{} as a function of the magnetic field strength $\gamma$.}
  \label{fig:lt_trans_-1_0_-2_0_quad}
\end{figure}

\begin{figure}[htbp]
  \centering
  \includegraphics*[scale=0.3]{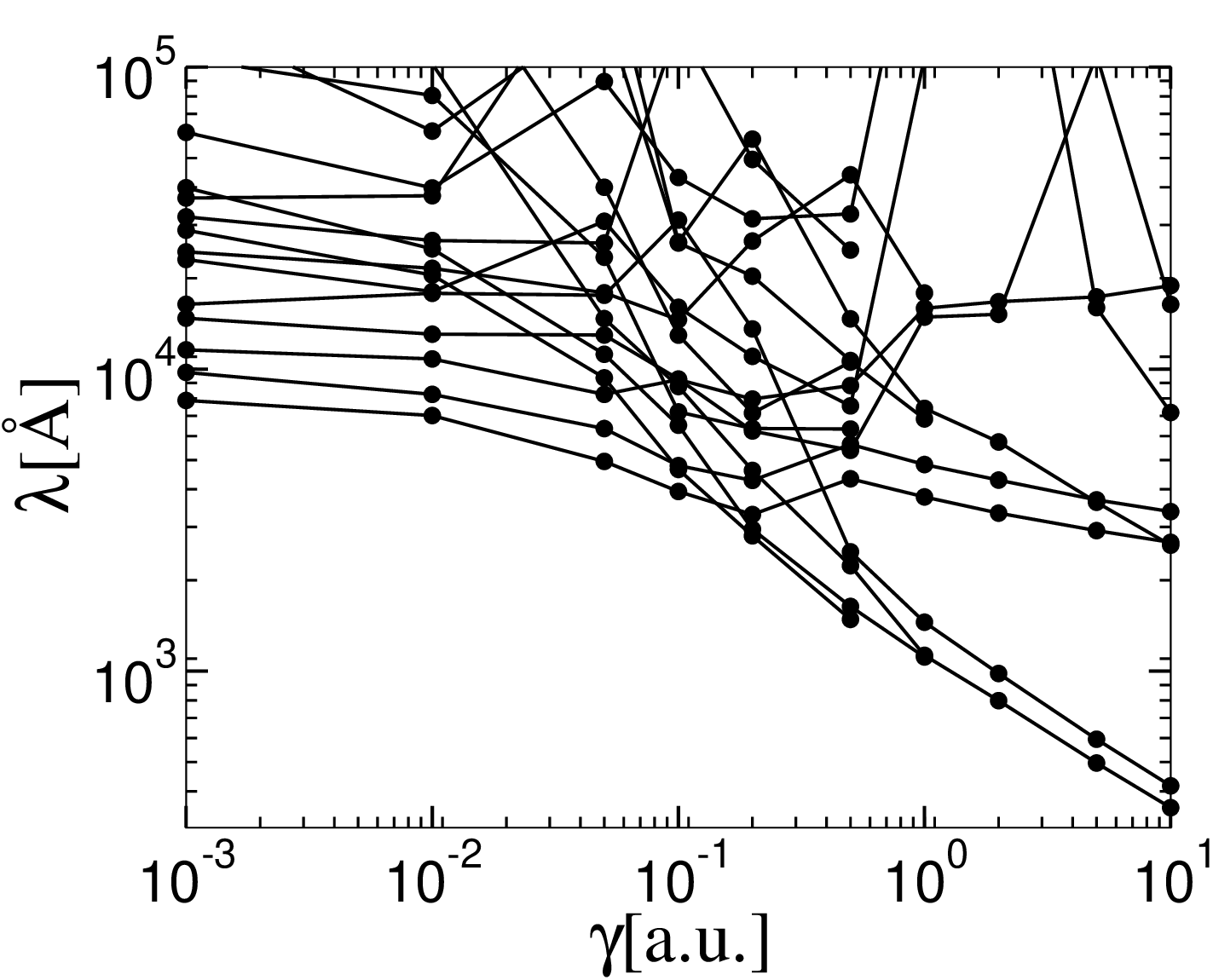}
  \caption{Transition wavelengths $\lambda$ for the circular polarized
  transitions $\nu^4(-2)^+\longrightarrow\mu^4(-3)^+$ ($\nu,\mu=1,2,3,4$)
  in \AA{} as a function of the magnetic field strength $\gamma$.}
  \label{fig:lt_trans_-2_0_-3_0}
\end{figure}

In this section we  present the results for the wavelengths $\lambda$
of the allowed 
electric dipole transitions. We will restrict the
wavelengths to the regime $\lambda<10^5$~\AA{} in order to avoid too
large uncertainties. In the
following we will 
consider the linear polarized transition
$\nu^4(-1)^+\longrightarrow\mu^4(-1)^-$ ($\nu,\mu=1,2,3,4$) (shown in
Fig.~\ref{fig:lt_trans_-1_0_-1_1}) and the circular polarized
transitions $\nu^{2S+1}M^+\longrightarrow\mu^{2S+1}(M-1)^-$
($\nu,\mu=1,2,3,4$) for ($M,S$)=($0,2$), ($-1,2$), ($-1,4$), and
($-2,4$) in Figs.~\ref{fig:lt_trans_0_0_-1_0} to \ref{fig:lt_trans_-2_0_-3_0}.

First, we  discuss some general features of the transition
wavelengths. For  the circular polarized transitions
($M,S$)=($0,2$), ($-1,2$), and ($-2,4$) (presented in
Figs.~\ref{fig:lt_trans_0_0_-1_0}, \ref{fig:lt_trans_-1_0_-2_0_dubl}, and
\ref{fig:lt_trans_-2_0_-3_0}) it can be observed, that some
transition wavelengths decrease in the limit of a strong field,
thereby following a power
law. Whereas for the linear polarized transition (Fig.~\ref{fig:lt_trans_-1_0_-1_1}), and the circular
polarized transition ($M,S$)=($-1,4$)
(Fig.~\ref{fig:lt_trans_-1_0_-2_0_quad}) such a behavior can not be
observed. The transitions, with the strongly decreasing wavelengths
are the ones, which involve triply tightly bound states. In our case  these are
the states
$1^20^+$, $1^2(-1)^+$, $1^2(-2)^+$ and $1^4(-3)^+$. The corresponding
wavelengths become for $\gamma=10$ shorter than $10^3$~\AA{}, whereas
the remaining 
transition wavelengths are in general longer than $10^3$~\AA.
In the symmetry subspaces, involved for the  linear
polarized transitions considered here, and the circular polarized transitions with
($M,S$)=($-1,4$), no triply tightly bound states exist.

For the linear polarized transitions
$\nu^4(-1)^+\longrightarrow\mu^4(-1)^-$ ($\nu,\mu=1,2,3,4$) shown in
Fig.~\ref{fig:lt_trans_-1_0_-1_1} it can be observed, that the
wavelengths in the low field regime ($\gamma<0.1$) are nearly
constant. In the   regime $0.1\leq\gamma\leq5$ the spectrum of wavelengths
becomes very complicated. This is due to  avoided crossings of excited
states being
present in both symmetry subspaces, that are involved in the
transitions. Especially we find divergences of the transition
wavelengths, being a consequence of crossovers of the energy levels.

In Fig.~\ref{fig:lt_trans_0_0_-1_0} the transition
wavelengths for the circular polarized transitions of the form
$\nu^2(0)^+\longrightarrow\mu^2(-1)^+$ ($\nu,\mu=1,2,3,4$) are shown.  In the
high field limit a bunch of 
small wavelengths, described above, can be identified easily. These
are transitions of the form $\nu^20^+\longrightarrow1^2(-1)^+$,
i.e. those involving the  state $1^2(-1)^+$. One
of these lines diverges at $\gamma\approx0.2$. It is associated with
the transition $1^20^+\longrightarrow1^2(-1)^+$. As mentioned above,
the energies of these two states become equal at
$\gamma=0.1929$. Further divergencies can be observed, which are caused by the
fact, that energy levels of the $M^{\Pi_z}=-1^+$
symmetry subspace increase much faster as a function of $\gamma$,
than those belonging to the 
$M^{\Pi_z}=0^+$ subspace.

For the case of the circular polarized doublet transitions
$\nu^2(-1)^+\longrightarrow\mu^2(-2)^+$ ($\nu,\mu=1,2,3,4$), displayed in
Fig.~\ref{fig:lt_trans_-1_0_-2_0_dubl}, the reader observes eight
lines, decreasing in the limit of strong fields thereby following a power
law. These lines correspond to the transitions involving the triply tightly
bound states $1^2(-1)^+$ and $1^2(-2)^+$. Furthermore the reader
should 
note, that the transitions among the other states show little variation, compared to the transitions
$\nu^20^+\longrightarrow\mu^2(-1)^+$, which is a consequence of 
the fact, that the energy levels in the participating symmetry
subspaces behave in a very similar way.

The corresponding quadruplet transitions
$\nu^4(-1)^+\longrightarrow\mu^4(-2)^+$ ($\nu,\mu=1,2,3,4$) depicted in
Fig.~\ref{fig:lt_trans_-1_0_-2_0_quad} follow a completely different
pattern. Because there are no triply tightly bound states in neither
of the subspaces, all the wavelengths are longer than $10^3$\AA{}. On
the other hand the behavior of the wavelengths in the  regime
$\gamma>0.1$ reflects  
the complicated energy level scheme of  both symmetry subspaces, which is
dominated by several avoided crossings. They result in energy
level crossovers and therefore devergencies for the corresponding
wavelengths.

For the transitions
$\nu^4(-2)^+\longrightarrow\mu^4(-3)^+$ ($\nu,\mu=1,2,3,4$) shown
in Fig.~\ref{fig:lt_trans_-2_0_-3_0} one observes some very short
transition wavelengths at $\gamma=10$ ($\lambda<400$~\AA),
that  correspond to 
transitions involving the high field ground state $1^4(-3)^+$, which
is a triply tightly bound state. For $\gamma>0.1$ avoided crossings
and the
rearrangement of energy levels creates a complex pattern.

\section{Summary and Conclusions}
\label{sec:concl}

In the present work we have investigated the electronic structure of the lithium
atom exposed to a strong homogeneous magnetic field. We cover the broad regime of 
field strengths $0 \le \gamma \le 10$ providing data for a grid of ten
values for the field strength.  
The key ingredient of our computational method is an anisotropic Gaussian basis set whose nonlinear variational
parameters (exponents!) are optimized for each field strength. These nonlinear optimizations, being based on a
sophisticated algorithmic procedure, are performed for {\it{one- and
    two-electron atomic systems}} in the
presence of the field. As a result we obtain a basis set of orbitals that allows for a rapidly
convergent numerical study of the electronic structure of, in particular, the lithium atom. 

Our computational approach to the three-electron problem is a full configuration interaction method.
This approach yields fully correlated wave functions that can, in principle, be determined to
arbitrary accuracy. To implement it for the above-mentioned basis set a number of techniques had to be
combined. To avoid linear dependencies of our non-orthogonal orbitals a cut-off technique with respect
to the overlap and Hamiltonian configuration matrix has been employed. Employing large configurational
basis sets of the order of several ten-thousand we arrived at relative accuracies of the order
of $10^{-4}$ for the total energies of the lithium atom in the presence of the field.

Total and one-particle ionization energies as well as transition wave lengths have been calculated
for the ground and typically three  excited states for each of the symmetries 
$^20^{+},^2(-1)^{+},^4(-1)^{+},^4(-1)^{-},^2(-2)^{+},^4(-2)^{+},^4(-3)^{+}$
thereby yielding a total of 28 states. This has to be compared with the
existing data on the lithium atom in the literature where only a few states for a few field strengths
have been investigated previously. Also, the predominant part of these investigations were not on a fully correlated level.

The ground state crossovers of the lithium atom with increasing field
strength were redetermined thereby yielding more precise values for
the crossover field strengths. A classification and discussion of the
one-electron ionization energies for the ground and excited states for
each of the above-given symmetries has been provided. Particular
emphasis has been given to the effects due to the tightly bound
orbitals and the singly or multiply tightly bound configurations. Only
a very limited number of states show a monotonically increasing
one-electron ionization energy in the complete regime of field
strengths considered here.  With increasing degree of excitation
avoided crossings lead to a nonmonotonous behavior of the energies.
For the electromagnetic transitions that involve states with tightly
bound orbitals we observe bundles of short wavelengths that decrease
monotonically with increasing field strength.

In principle our approach allows investigations of more than three
electron atoms. Furthermore since it yields the eigenfunctions
arbitrary properties and in particular oscillator
strengths for lithium and  more electron atoms can be obtained.


\end{document}